\documentclass[10pt,a4paper]{article}
\usepackage{authblk}
 \usepackage[utf8]{inputenc}
 \usepackage{amsmath}
 \usepackage{amsfonts}
 \usepackage{amssymb}
 \usepackage{makeidx}
 \usepackage{graphicx}
 \usepackage{lmodern}
 \usepackage{kpfonts}
\numberwithin{equation}{section}
\usepackage{fullpage}
\usepackage{cite}
\usepackage[T1]{fontenc}
\usepackage{mathtools}
\newcounter{mysubequations}
\usepackage{pict2e}

\makeatletter
\DeclareRobustCommand{\loplus}{\mathbin{\mathpalette\dog@lsemi{+}}}
\DeclareRobustCommand{\lotimes}{\mathbin{\mathpalette\dog@lsemi{\times}}}
\DeclareRobustCommand{\roplus}{\mathbin{\mathpalette\dog@rsemi{+}}}
\DeclareRobustCommand{\rotimes}{\mathbin{\mathpalette\dog@rsemi{\times}}}

\newcommand{\dog@rsemi}[2]{\dog@semi{#1}{#2}{-90,90}}
\newcommand{\dog@lsemi}[2]{\dog@semi{#1}{#2}{270,90}}
\newcommand{\dog@semi}[3]{%
  \begingroup
  \sbox\z@{$\m@th#1#2$}%
  \setlength{\unitlength}{\dimexpr\ht\z@+\dp\z@\relax}%
  \makebox[\wd\z@]{\raisebox{-\dp\z@}{%
    \begin{picture}(1,1)
    \linethickness{\variable@rule{#1}}
    \roundcap
    \put(0.5,0.5){\makebox(0,0){\raisebox{\dp\z@}{$\m@th#1#2$}}}
    \put(0.5,0.5){\arc[#3]{0.5}}
    \end{picture}%
  }}%
  \endgroup
}
\newcommand{\variable@rule}[1]{%
  \fontdimen8  
  \ifx#1\displaystyle\textfont3\else
    \ifx#1\textstyle\textfont3\else
      \ifx#1\scriptstyle\scriptfont3\else
        \scriptscriptfont3\relax
  \fi\fi\fi
}
\makeatother

\newcounter{my-counter} 

\newcounter{subsubsubsection}[subsubsection]
\def\subsubsubsectionmark#1{}

\def\subsubsubsection{\@startsection
     {subsubsubsection}{4}{\z@} {-3.25ex plus -1
     ex minus -.2ex}{1.5ex plus .2ex}{\normalsize\bf}}
\def\l@subsubsubsection{\@dottedtocline{4}{4.8em}
     {4.2em}}

\title{ Time evolution operator for a  $h(1) \oplus h(1) \roplus  u(2)$ time-dependent quantum Hamiltonian;  
a self-consistent resolution method based on Feynman's disentangling rules }

 \author{Nibaldo Edmundo Alvarez Moraga\footnote{Email address: nibaldo.alvarez.m@mail.pucv.cl} }
 
  \affil{ Autonomous Center of Theoretical Physics and Applied Mathematics, \\
2805 Place de Darlington, Montreal (Quebec), H3S1L4, Canada}

\begin{document}

   \maketitle

\begin{abstract}
 In this article the time evolution operator of two interacting quantum oscillators, whose Hamiltonian is an element of the complex $\left\{ h(1) \oplus h(1) \right\} \roplus u(2)$ algebra,  is analyzed using the Feynman time ordering operator  techniques. This method is consistently  used  to  generate the conditions and to formally find explicit disentangled expressions for such   
operator. In this way, it is shown  that  all the problem reduces to solve a complex Riccati-type differential equation. Some closed solutions to this differential equation are found and then concrete disentangling expressions for the time-ordered evolution operator are given. Finally, the time evolution of the coherent states linked to the isotropic 2D quantum oscillator are analyzed under alternative time-independent an time-dependent Hamiltonian systems.

 \end{abstract}
 
\section{Introduction}
\label{sec-zero}
In quantum mechanics the time evolution of the state of a system is determined, depending on the formulation we are working with, by solving the Schrödinger equation,  by constructing a unitary time evolution operator or by determining the so-called propagator\cite{Schro}--\cite{RF-AH}. Certainly, all these formulations are equivalent and the use of one or the other to describe the behavior or a certain property of a quantum system will give the same results, the only difference could be in the fact of weighting in which of them the calculation of a certain property of the system we are studying is easier to realize, although there is no single answer here either. 

 In the present article, we will disentangle the unitary time evolution operator which associated time-dependent Hamiltonian is an element of the complex Lie algebra $\left\{ h(1) \oplus h(1)\right\} \roplus u( 2) .$ This algebra is a composed algebra formed as the semi-direct sum of the direct sum of two Heisenberg-Weyl Lie algebras $h(1)$ and the oscillator algebra $u(2).$ The Hamiltonian this way constructed represents, in the two boson representation space, two interacting oscillators with linear external coupling added. In this sense, this article is a special case of the case of  $N$ interacting oscillators already treated in the literature \cite{KuMe}. However, here we will use a self-consistent technique to disentangle completely the time evolution operator, that is, the Feynman's operator disentangling rules \cite{Feyn-51},  and we will construct some explicit expressions of it.
 
 This article is organized  as follows :  In section \ref{sec-one} we briefly recall the concept of time evolution of a physical state of a quantum system based on the  action of a time  ordered unitary operator on the initial state of such a system and at the same time we present the Feynman's disentangling  operator technique, which allows to isolate  an exponential operator from a unitary operator of the time evolutiontype. There we also show the equivalence between this technique and that based on the application of a unitary similarity transformation. In section \ref{sec-two},  we begin by expressing the time evolution operator as the product of two exponential factors, the first having the structure of a unitary displacement operator of the two dimensional quantum  oscillatortype, which acts absolutely at right of the second, which in turn has the structure of a unitary time evolution  operator whose Hamiltonian belongs to the algebra $u(2).$ Then, we proceed to disentangle the second exponential factor and establish the conditions under which a complete disentangling of the time evolution operator  is possible. These conditions assume the form of a complex Riccati-type differential equation. In section \ref{sec-three} a closed solution of the complex Riccatitype equation for special choices of the time dependent parameters is given, some of them suggested by the form of the standard  structure  of a disentangled   time-independent element of the $SU(2)$ group. In section \ref{sec-four} we compute the time evolution of a class of coherent states linked to  the    2 D isotropic harmonic oscillator\cite{JM-VH}. We show that these states are also linked to a class of time-independent $u(2)$ Hamiltonian  that can  be reduced by means of a unitary squeezing  mixed operator to a  one mode harmonic oscillator. In section \ref{sec-five}, we  repeat  that we did in the previous section but now with a class  of  coherent states linked to  a  2 D  time-dependent $u(2)$-type of  interacting oscillators. In  section \ref{sec-six} we try another option of  disentangling ordering for the unitary time evolution operator. There, we show the equivalence of the resulting equations with those of the previous choice. Finally, in appendix \ref{appa}, by choosing a suitable representation space, we transform the problem of solving a  system of linear differential equations into the equivalent one of solving a time ordered non-homogeneous Schr\"odinger-type equation\cite{NEAM-92}.   This last fact will allow us to choose the $j=\frac{1}{2} $ representation space of the $su(2)$ algebra to compute the  matrix elements which are necessary to finish  the  construction  of the  above mentioned  two-dimensional displacement unitary operator.

\section{Unitary time evolution operator}
\label{sec-one}
In quantum mechanics, the connection between theory and experiment is given by the expectation values of the dynamical variables. For a system represented by the Hamiltonian $\hat{H} (t),$  if  $\hat{\Lambda}$ denotes a  dynamical variable, its expectation value is given by

\begin{equation}
\langle \Lambda \rangle (t) = \langle \Psi (t) \mid \hat{\Lambda} \mid \Psi (t) \rangle,
\end{equation}
where the state $\mid \Psi (t) \rangle $ is determined by  the Schr\"odinger equation

\begin{equation}
i  \hbar \frac{d}{dt} \mid \Psi (t) \rangle = \hat{H} (t)  \mid \Psi (t) \rangle \label{eq-schrodinger}
\end{equation}
and the initial condition, when $t=t_0,$ $\mid | \psi (t_0) \rangle.$ 
The solution of (\ref{eq-schrodinger}) can be written in the form

\begin{equation}
\mid \Psi (t) \rangle = \hat{U} (t,t_0) \mid \Psi (t_0) \rangle, 
\end{equation}
where $ \hat{U} (t,t_0) $ is the time-ordered unitary operator  of the system which verifies the differential equation

\begin{equation}
i \hbar \frac{d}{dt}  \hat{U} (t,t_0) = \hat{H}  \hat{U} (t,t_0), \label{eq-schrodinger-U}
\end{equation}
with the initial condition $ \hat{U} (t_0,t_0)=\hat{I}.$
The solution of (\ref{eq-schrodinger-U}) can be formally written as

\begin{equation}
\hat{U} (t,t_0) = e^{ - \; \frac{i }{\hbar} \int_{t_0}^{t}  \left[ \hat{H} (s) \right]_{+} ds},  \label{unitary-U}
\end{equation} 
where the subscript $+$ in the integral factor indicates that, when expanding the exponential operator, the order of action of the Hamiltonian $\hat{H} (t)$ on the states depends on the value of the parameter $s,$  that is, the Hamiltonian evaluated at an earlier time acts first than a Hamiltonian evaluated at a later time.
There are several important properties this operator verifies,  among them we have

\begin{equation}
\hat{U} (t, 0) =   \hat{U} (t, t_0)  \hat{U} (t_0, 0), \quad {\rm which \;  is \;  equivalent \;  to} \quad  \hat{U} (t, t_0) =   \hat{U} (t, 0)  \hat{U}^{-1} (t_0, 0), \label{U-property}
\end{equation}     
where $t \geq t_0 \geq 0.$

\subsection{Feynman's disentangling rule and unitary transformations}
In this section we use the  Feynman's  disentangling rule to factorize a the time-ordered unitary operator in terms of exponential factors.  To do that,  let us  start with any  generic  time-ordered operator of the form

\begin{equation}
\hat{V} (t,t_0) =  e^{- \frac{i}{\hbar} \int_{t_0}^{t}  \left[\hat{W} (s) \right]_+  ds  }.
\end{equation}
Let us now rewrite this operator in the form

\begin{equation}
\hat{V} (t,t_0) =  e^{- \frac{i}{\hbar} \int_{t_0}^{t}  \left[\hat{W} (s) -  i \hbar \hat{\gamma} (s) +  i \hbar \gamma (s) \right]_+  ds  },
\end{equation}
where we only have added to $\hat{W} (s)$ the null operator   $ i  \hat{\gamma} (s) - i \hat{\gamma} (s).     $
Now, applying the Feynman's  disentangling rule to this last operator we get

\begin{equation}
\hat{V} (t,t_0) = \hat{S}_{t}     e^{- \frac{i}{\hbar} \int_{t_0}^{t} \left[  \hat{S}^{-1}_s  \left(\hat{W} (s)  -  i \hbar \gamma (s) \right) \hat{S}_s \right]_{+}  ds  }  \;  \hat{S}^{-1}_{t_0}      , \label{V-time-ordered}
\end{equation}
where the $ \hat{S}_{s}  $ operator verifies

\begin{equation}
\frac{d \hat{S}_s   }{ds}  = \hat{\gamma} (s) \hat{S}_s, \label{S-time-ordered}
\end{equation}
i.e., $\hat{S}_{s}$ is the time-ordered operator
\begin{equation}
\hat{S}_s = e^{\int_{t_0}^{s} [\hat{\gamma} (\tau) ]_+   d\tau} \; \hat{S}_{t_0}
\end{equation}
 Finally, by combining equations (\ref{V-time-ordered} ) and  (\ref{S-time-ordered}), we obtain

\begin{equation}
\hat{V} (t,t_0) = \hat{S}_{t}     e^{- \frac{i}{\hbar} \int_{t_0}^{t} \left[  \hat{S}^{-1}_s  \hat{W} (s) \hat{S}_s  +  i   \hbar    \left( \frac{d }{ds}\hat{S}^{-1}_{s}  \right) \hat{S}_s \ \right]_{+}  ds  } \; \hat{S}_{t_0}. \label{V-time-ordered-final}
\end{equation}
 We note that if $i \hat{\gamma}_s$ is Hermitian then $\hat{ S}_s$ is unitary, that is, ${\hat S}_s^{-1} = {\hat S}_s ^ {\dagger}.$ The expressions (\ref{V-time-ordered}) and (\ref{V-time-ordered-final}) are certainly equivalent, but while the former is more useful for  practical purposes,    to generate conditions under which a determined exponential operator can be factorized, for example, the   latter is more indicated at the moment of analyzing  the general structure of the quantum system we are studying.

\section{Time evolution operator for the time-dependent $h(1) \oplus h(1) \roplus u(2)$ Hamiltonian}
\label{sec-two}
The starting point of the present study is the two mode time-dependent Hamiltonian

\begin{equation}
\hat{H} (t) = \sum_{\sigma, \lambda=1}^{2}  \hbar  w_{\sigma \lambda} (t) \hat{a}^\dagger_{\sigma} \hat{a}_{\lambda} +  \sum_{\sigma=1}^{2} [ F_{\sigma} (t) \hat{a}^\dagger_{\sigma} +  F^\ast_{\sigma} (t)  \hat{a}_{\sigma} ] + B(t) \hat{I},  \label{hamiltonian-two-modes}
\end{equation}
where we have used essentially  the same notation and structure used by S. Kumar and C. L. Mehta in \cite{KuMe}, but now with the restriction that the $\sigma$ and $\lambda$ indexes   run only from 1 to 2.  As $\hat{H} (t)$ is Hermitian, $ w_{11} (t), w_{22} (t)$ and $B (t)$ are real functions of time  while   $w_{12} (t) = w_{21}^\ast (t)$ and  $F_{\sigma} (t)$ are arbitrary complex functions of time. The set of creation and annihilation operators satisfy the $h(1) \oplus h(1)$ commutation relations

\begin{equation}
[\hat{a}_\sigma,\hat{a}_\lambda] = [\hat{a}^\dagger_\sigma,\hat{a}^\dagger_\lambda] = 0, \quad  [\hat{a}_\sigma,\hat{a}^\dagger_\lambda] = \delta_{\sigma \lambda}, \quad \sigma,\lambda=1,2. \label{commutation-rel}
\end{equation}
The time evolution operator for the Hamiltonian (\ref{hamiltonian-two-modes}) formally writes

\begin{equation}
\hat{U} (t,t_0) = \exp{\left[ - \;  \frac{i}{\hbar}  \; \int_{t_0}^{t}  \left[\sum_{\sigma, \lambda=1}^{2}  \hbar  w_{\sigma \lambda} (s) \; \hat{a}^\dagger_{\sigma} \hat{a}_{\lambda} +  \sum_{\sigma=1}^{2} \; [ F_{\sigma} (s) \; \hat{a}^\dagger_{\sigma} +  F^\ast_{\sigma} (s) \;  \hat{a}_{\sigma} ] + \beta (s) \right]_+ ds \right]}. \label{U-starting-non-disentangled}
\end{equation}
This operator have  already been formally disentangled  in the case  of arbitrary $N$ interacting quantum harmonic oscillators \cite{KuMe} by means of two successive  unitary similarity transformations, the first of translation and the second of rotation. There, some conditions on the time-dependent $w_{\sigma \rho} (t) $ parameters have been given for the time evolution could be disentangled.   Here, we will disentangle this operator for the case $N=2$ by isolating an exponential operator from the time evolution operator, with the help of the  Feynman's  time ordered operator technique  \cite{Feyn-51}. It has been shown in the literature that these two techniques are completely equivalent \cite{NEAM-92}, they lead exactly to the same  results, but Feynman’s technique gives us more flexibility in analyzing the conditions that allow us to solve the differential equations necessary to obtain an explicit disentangled expression  for the time evolution operator.  

 \subsection{Disentangling with the help of a unitary displacement operator}
In this section we factorize the two mode time evolution operator with the help of a unitary  displacement operator. In accordance with equation (\ref{U-property}) we only need to compute $\hat{U} (t,0).$ Thus, by using the Feynman disentangling rules, we can write this time evolution operator in the  form
 
 \begin{eqnarray}
 \hat{U} (t,0) & =& \hat{S} (t)  \exp \left[ - \;  \frac{i}{\hbar}  \; \int_{0}^{t} \hat{S}^{-1} (s)  \left[\sum_{\sigma, \lambda=1}^{2}  \hbar  w_{\sigma \lambda} (s) \; \hat{a}^\dagger_{\sigma} \hat{a}_{\lambda}  \right.  \right. \nonumber \\  &+ &
 \left.  \left.  \sum_{\sigma=1}^{2} \; [ ( F_{\sigma} (s)  - i \hbar \gamma_{\sigma} (s)) \; \hat{a}^\dagger_{\sigma} +  ( F^\ast_{\sigma} (s)  + i \hbar \gamma^\ast_{\sigma} (s ) )\;  \hat{a}_{\sigma} ] + \beta (s) \right] \hat{S} (s) ds \right]_+ ,  \label{U-time-evolution-t-0}
\end{eqnarray}
 where 
 \begin{equation}
 \hat{S} (s) = \exp{  \int_{0}^{s}  \sum_{\sigma=1}^{2} [\gamma_\sigma (\tau) \hat{a}^{\dagger}_{\sigma}  - \gamma^\ast_\sigma (\tau) \hat{a}_{\sigma}   ]_+ d\tau },
 \end{equation}
with the choice $\hat{S} (0) =\hat{I}.$ 
The unitary displacement  operator $\hat{S} (s)$ is by itself an exponential operator that need to be disentangled for operating on the states or for combining with the other operators. Then, its disentangled form can be reached by using again the Feyman factorization rules:

\begin{equation}
\hat{S} (s) = \hat{V} (s) \exp{\left[   -  \int_{0}^{s}  \sum_{\sigma=1}^{2}   \hat{V}^{-1} (\tau)     \gamma^\ast_\sigma (\tau) \hat{a}_{\sigma}  \hat{V} (\tau)  d\tau \right]} \label{S-unitary}
\end{equation}
where

\begin{equation} 
\hat{V} (\tau) =  \exp{ \left[ \int_{0}^{\tau}  \sum_{\sigma=1}^{2} [\gamma_\sigma (\nu) \hat{a}^{\dagger}_{\sigma} d\nu \right]}, \label{V-unitary}
\end{equation}
with the choice $\hat{V} (0) = \hat{I}.$ We note that according to the commutation relations (\ref{commutation-rel}),  no subscript $+$ is needed in (\ref{S-unitary}) and  (\ref{V-unitary}).  Indeed, 

\begin{equation}
\hat{V} (\tau) =  \exp{ \left[ \sum_{\sigma=1}^{2}  c_{\sigma} (\tau )  \hat{a}^{\dagger}_{\sigma} \right]}, \label{V-unitary-explicit}
\end{equation}
where $c_{\sigma} (\tau) = \int_{0}^{\tau}  \gamma_{\sigma} (\nu) d\nu,$ or  $\frac{d}{d\tau} c_{\sigma} (\tau) = \gamma_{\sigma} (\tau).  $ Thus, by inserting (\ref{V-unitary-explicit}) into (\ref{S-unitary}) and using the relation $ e^{-  c_{\mu} (\tau) \hat{a}^\dagger}  \hat{a}_{\sigma}   e^{c_{\mu} (\tau) \hat{a}^\dagger} = \hat{a}_{\sigma} +  c_{\sigma}  (\tau)  \delta_{\sigma \mu}, $ we get   

\begin{equation}
\hat{S} (s) =  \exp{ \left[ \sum_{\sigma=1}^{2}  c_{\sigma} (s )  \hat{a}^{\dagger}_{\sigma} \right]}  \exp{\left[  -  \sum_{\sigma=1}^{2}         c^\ast_\sigma (s)  \hat{a}_{\sigma}   -  \sum_{\sigma=1}^{2} \int_{0}^{s}  \gamma^\ast_{\sigma} (\tau) c_{\sigma} (\tau) d\tau    \right]},
\end{equation}
which after some manipulation becomes

\begin{equation}
\hat{S} (s) =   \exp\left[ \sum_{\sigma=1}^{2}  [ c_{\sigma} (s )  \hat{a}^{\dagger}_{\sigma}   -    c^\ast_\sigma (s)  \hat{a}_{\sigma}] \right] 
\exp\left[  \frac{1}{2}  \sum_{\sigma=1}^{2} \int_{0}^{s}  [ \gamma_{\sigma} (\tau)  c^\ast_{\sigma} (\tau)  - \gamma^\ast_{\sigma} (\tau) c_{\sigma} (\tau) ] d\tau   \right]. 
\end{equation}
Inserting this last result into  equation (\ref{U-time-evolution-t-0}), then using the generic displacement operator property $\hat{D}^\dagger  (c_{\mu} (s)) \hat{a}_{\sigma}  \hat{D} (c_{\mu} (s)) = \hat{a}_\sigma + c_{\sigma} (s) \delta_{\sigma \mu,}$ and its corresponding conjugate relation,  and requiring that the linear terms in the creation and annihilation  operators under the integral sign  be eliminated from the exponential factor, we get 

\begin{eqnarray}
 \hat{U} (t,0) &=& \exp\left[ \sum_{\sigma=1}^{2}  [ c_{\sigma} (s )  \hat{a}^{\dagger}_{\sigma}   -    c^\ast_\sigma (s)  \hat{a}_{\sigma}] \right]  \nonumber \\ & \times &
 \exp \left[ - \;  \frac{i}{\hbar}  \; \int_{0}^{t}    \left[\sum_{\sigma, \lambda=1}^{2}  \hbar  w_{\sigma \lambda} (s) \; \hat{a}^\dagger_{\sigma} \hat{a}_{\lambda} +  \frac{1}{2}    \sum_{\sigma=1}^{2} \; [ ( F_{\sigma} (s)  c^\ast_{\sigma} (s))  +   F^\ast_{\sigma} (s)   c_{\sigma} (s ) )]
 + \beta (s) \right]   ds \right]_+ ,
\end{eqnarray}
where

\begin{equation}
i \hbar \frac{d c_{\sigma} (t)}{dt} - \sum_{\lambda =1}^2 \hbar w_{\sigma \lambda} (t) \;  c_{\lambda} (t) - F_{\sigma} (t) =0, \quad \sigma=1,2.  \label{eq-sys-differential}
\end{equation}
 By the way, these equations are the same found in \cite{KuMe} with a little difference in the sign of  $c_{\sigma} (t), \sigma=1,2.$ 
 In the appendix \ref{appa},  we use the time order operator method to obtain a formal set of solutions for this coupled differential equations\cite{NEAM-92}. Explicit expressions of these solutions are given in equation (\ref{c-solutions}).  The advantage of the method detailed there lies in the fact that one can choose at will the space of representation of the $u(2)$ generators.  As we will see below, in our case, a  suitable irreducible representation is the standard $\mid j, m \rangle$ space,  where $j=\frac{1}{2},$ and $m= \pm \frac{1}{2}.$ On the other hand, we will see in the next section, the same solutions will  allow  us to disentangle the remaining $u(2)$ part of the total unitary time evolution operator. This last detail have been already  been pointed out by  S. Kumar and C. L. Mehta in  \cite{KuMe}.
 
 \subsection{ Two boson realization of the $u(2)$ generators} 
The remaining unitary operator we need to disentangle is quadratic in the creation and annihilation operators, that  is,

\begin{equation} \hat{U}_0  (t,0) = \exp \left[ - \;  \frac{i}{\hbar}  \; \int_{0}^{t}    \left[\sum_{\sigma, \lambda=1}^{2}  \hbar  w_{\sigma \lambda} (s) \; \hat{a}^\dagger_{\sigma} \hat{a}_{\lambda}  \right]   ds \right]_+. \label{second-Unitary-operator}
\end{equation}
Here, it is convenient to express it in terms of the 
 Schwinger two boson realization \cite{Schwinger} of the $su(2)$ Lie algebra generators\footnote{In this article, we write the two-modes creation and annihilation operators in the following  simplified form: $\hat{a}_1 \equiv \hat{a}_1 \otimes \hat{I}_2,$   $ \hat{a}_2 \equiv \hat{I}_1 \otimes \hat{a}_2 ,$ $ \hat{a}_{1}^\dagger \equiv \hat{a}_{1}^\dagger \otimes \hat{I}_2 $ and  $ \hat{a}_2^\dagger \equiv \hat {I}_1 \otimes \hat{a}_2^\dagger,$ then, par example, $ \hat{a}_1^\dagger \hat{a}_2 =  (\hat{a}_1^\dagger \otimes \hat{I}_2 ) (\hat{I}_1 \otimes \hat{a}_2   ) = \hat{a}_1^\dagger \otimes \hat{a}_2. $ Sometimes,   we will also write $ \hat{a}_2^\dagger \hat{a}_1$  in place of  $\hat{a}_1 \hat{a}_2^\dagger =   \hat{a}_1  \otimes \hat{a}_2^\dagger,$ for example. }:
\begin{equation}
\hat{J}_+ = \hat{a}_1^\dagger \hat{a}_2, \quad \hat{J}_- = \hat{a}_1 \hat{a}_2^\dagger,  \quad  \hat{J}_3 =  \frac{1}{2} (\hat{a}_1^\dagger \hat{a}_1 - \hat{a}_2^\dagger \hat{a}_2 ),    
   \end{equation}
which verify the commutation relations

\begin{equation}
[\hat{J}_3, \hat{J}_\pm]= \pm \; \hat{J}_\pm, \quad [\hat{J}_{+}, \hat{J}_{-}]= 2 \hat{J}_{3},   
  \end{equation} 
 and the Casimir operator 
\begin{equation} 
 \hat{N} =  \frac{1}{2} (\hat{a}_1^\dagger \hat{a}_1 + \hat{a}_2^\dagger \hat{a}_2 ),    
\end{equation}
 which commutes with all them.
Thus the unitary operator assumes the form

\begin{equation} \hat{U}_0  (t,0) = \exp \left[ - \;  i \; \int_{0}^{t}  \left[ \frac{d\alpha(s) }{ds}  \hat{N}  +  \frac{ d\rho (s)}{ds} \hat{J}_3 + w_{12} (s) \hat{J}_+   + w_{21} (s) \hat{J}_-   \right]_+                                    ds  \right] \label{U-t-0-expression}
\end{equation}
where we have defined the auxiliary functions $\alpha(s)$ and $\rho(s)$ in such a way that $ \frac{d\alpha (s)}{ds} = w_{11} (s) + w_{22} (s)$ and $\frac{d\rho (s)}{ds}= w_{11} (s) - w_{22} (s),$ with $\alpha(0)=\rho(0)=0.$

 At this stage, to untangle (\ref{U-t-0-expression}) we have several choices.  All these choices are completely equivalent with respect to the complexity of the differential equations that we finally have to solve to obtain the final untangled form of the unitary time evolution operator.   In fact, once such a unitary operator has been completely untangled following a determined choice, we can move from one to another simply by using appropriate similarity transformations formed from the untangled exponential factors.

\subsubsection{A simple and direct choice of disentangling }
For example,  we could start by isolating an exponential factor raised to a factor proportional to $\hat{N},$ followed by another exponential factor raised to a factor proportional to  $\hat{J}_3,$ to finally  disentangle the remaining exponential factor raised to a given linear combination of the ladder operators $\hat{J}_{\pm}.$ As $\hat{N}$ commutes with all $su(2)$ generators and $e^{i \rho (s) \hat{J}_3} \hat{J}_{\pm} e^{-i \rho (s) \hat{J}_3} = e^{\pm  i \rho (s)} \; \hat{J}_\pm,$ a  reiterated application of the Feynman disentangle  exponential factor`s rule leads without any difficulty to

\begin{equation} \hat{U}_0  (t,0) =  e^{ -i \alpha (t) \hat{N}}  \; e^{-i \rho (t) \hat{J}_3}  
\exp \left[  \; \int_{0}^{t}  \left[   \eta (s) \hat{J}_+  -  \eta^\ast (s) \hat{J}_-   \right]_+   ds  \right],
\end{equation}
where $\alpha(s)= \int_{0}^{s} [w_{11} (s) + w_{22} (s) ] ds, $   $\rho (s)= \int_{0}^{s} [w_{11} (s) -  w_{22} (s) ] ds $  and $\eta (s) = - i w_{12} (s) e^{i \rho (s)}.$ 
 
The disentangling of the last factor
\begin{equation}
\hat{T} (t) = \exp\left[ \int_{0}^{t}  \left[   \eta (s) \hat{J}_+  -  \eta^\ast (s)  \hat{J}_-   \right]_+   ds  \right],  \label{T-unitary-operator} 
   \end{equation}
is not always possible, it depend on the form of the functions $w_{12} (s)= w^\ast_{21} (s)$ and of the real functions
$w_{11} (s)$ and $w_{22}( s).$   Indeed,  a first factorization in terms of exponential operators is

\begin{equation}
\hat{T} (t) =  e^{\Lambda(t) \hat{J}_+}
\exp \left[  \int_{0}^{t}   e^{- \Lambda (s) \hat{J}_{+}  } \left[   (\eta (s)  -  \lambda(s) ) \hat{J}_+  - \eta^\ast (s) \hat{J}_-   \right] e^{ \Lambda (s) J_+}   ds  \right] , \label{T-su2}
 \end{equation}
 where $\Lambda (s) = \int_{0}^{s} \lambda (\tau) d\tau $ or $\lambda (s) = \frac{d\Lambda (s)}{ds}.$ By using in (\ref{T-su2}) the relation $e^{ - \Lambda (s) \hat{J}_+} \; \hat{J}_-  \; e^{  \Lambda (s) \hat{J}_+} = \hat{J}_{-} - 2 \Lambda(s) \hat{J}_3 - (\Lambda (s))^2 \hat{J}_+$ and making a suitable regrouping of the terms we get

\begin{equation}
\hat{T} (t) =  e^{\Lambda(t) \hat{J}_+}
\exp \left\{ \int_{0}^{t}   \left[ [ \eta(s)  -  \lambda(s) +  \eta^{\ast} (s) (\Lambda (s))^2 ] \hat{J}_+   - \eta^{\ast} (s) [ \hat{J}_{-}  -  2 \Lambda (s)  \hat{J}_3 ] \right]_+   ds  \right\} . \label{T-su2-dis-1}
 \end{equation}
 By  forcing  the coefficient  of $\hat{J}_{+}$ in the second exponential factor to be null, we get 
 
\begin{equation}
\hat{T} (t) =  e^{\Lambda(t) \hat{J}_+}
\exp \left\{ - \int_{0}^{t}   \;  \eta^{\ast} (s) \;   \left[ \hat{J}_{-}  -  2  \Lambda (s) \hat{J}_3 ] \right]_+   ds  \right\} . \label{T-su2-dis-2}
 \end{equation}
 with the condition 
 
 \begin{equation}
 \frac{d\Lambda (s)}{ds}  - \eta (s) - \eta^\ast (s) (\Lambda (s))^2 =0,  \quad \Lambda (0) =0, \label{eq-Riccati}
\end{equation}  
 which a differential equation of the Riccati type. 
 
 Finally, by performing the disentangling of  the remaining exponential terms we get
  
  \begin{equation}
\hat{T} (t) =  e^{\Lambda(t) \hat{J}_+}  e^{\Omega (t) \hat{J}_{3} } e^{\Gamma (t) \hat{J}_{-} }, \label{T-su2-dis-3}
 \end{equation}
where 

\begin{equation}
\Omega (t) =  2   \int_{0}^{t} \eta^{\ast} (s) \Lambda (s) \; ds \quad \text{ and}  \quad \Gamma (t) = -   \int_{0}^{t} \eta^{\ast} (s)  e^{\Omega (s)} ds.  \label{Omega-Gamma-expressions}
  \end{equation}

\section{A closed solution of the complex Riccati type equation}
\label{sec-three} 
It is clear if we  were  able to solve  equation (\ref{eq-Riccati})  for any arbitrary  value of the time dependent functions $w_{\sigma \lambda} (s)$  we would have succeed in our task, but a closed solution of this differential is not always possible. For a formal discussion  about it see \cite{N-Steinmetz}--\cite{MHSAM}, and  references therein, for example. Here we will generate some closed solutions of this equation by appealing to some structural criteria of the operator algebra. 

\subsection{Standard $su(2)$ type structure}

   A first good criterion to consider is the unitary character the operator $\hat{T}$ and its counterbalanced  exponential structure in terms of the $su(2)$ generators. Then, similarly  to the case when all coefficients are time-independent constants, we can choose  $\Omega (t)$ to be  real at any time, which  implies that, as we will see in the following  development,   $\Gamma (t) = - \Lambda^\ast (t).$  
A sufficient condition for   $\Omega (t)$ to be real is

\begin{equation}
\eta^{\ast} (s)  \Lambda (s) = \eta (s) \Lambda^\ast (s)\label{sufficient-condition}.
\end{equation}

Taking into account  this condition in equation (\ref{eq-Riccati}) we get
\begin{equation}
 \frac{d\Lambda (s)}{ds}  =  \eta (s) (1 +  \| \Lambda (s) \|^2 ).  \label{eq-Riccati-condition}
\end{equation} 
 Now, if in this last equation we use the polar forms $\Lambda (s) = R(s) e^{i \theta(s)}$ and $\eta (s) = \| \eta (s) \| e^{i \phi (s)} $, where $R (s), \theta (s)$ and $\phi (s)$ are real functions, and we separate it into its real and imaginary parts, we get
 
 \begin{equation}
 \frac{dR (s)}{ds} =  (-1)^k \;  \| \eta (s) \|  (1 + R^2 (s)), \quad k = 0, \pm 1, \cdots, \quad \label{eq-R}
 \end{equation}
and the condition $\theta (s) = \theta_{0}, \forall s $ and $\phi (s) = \phi_0 = \theta_0 - k \pi, \; k=0, \pm 1, \cdots, \; \forall s, $ where $\phi_0$ and $ \theta_0 $ are real constants. 
Thus  by  integrating equation (\ref{eq-R}) we obtain

\begin{equation}
R (s) =  (-1)^k \;  \tan [ \int_{0}^{s}  \| \eta (\tau) \| d\tau ], \quad k=0, \pm 1, \cdots. 
\end{equation}
 Finally, taking into account all these results, we find a  solution of equation ( \ref{eq-Riccati}), that is,
 
  \begin{equation}
 \Lambda (s) =  \tan [\int_{0}^{s} \|  \eta (\tau) \|  d\tau  ] \;  e^{i \phi_0 },  \label{sol-Riccati}
 \end{equation}
provided that 
\begin{equation}
\eta (s) = \| \eta (s) \| \; e^{i \phi_0} . \label{condition-phase-constant}
\end{equation}
 Hence, the $\Omega (t)$ function writes
 
 \begin{equation}
 \Omega (t) =  2   \int_{0}^{t}  \| \eta (s) \|  \tan [ \int_{0}^{\tau} \|  \eta (\tau) \| \; d\tau] ds= 2 \ln \left(  \sec  [\int_{0}^{s} \| \eta (\tau) \| d\tau  ]   \right)
 \end{equation}
and the  $ \Gamma (t)$ function assumes the form

\begin{equation}
 \Gamma (t) =  - \int_{0}^{t} \| \eta (s) \|  e^{-i \phi_0}  \sec^2  [\int_{0}^{t} \|  \eta (\tau) \|  d\tau  ]   ds = -  \;  e^{-i \phi_0} \tan[\int_{0}^{t} \| \eta (s) \|  \; ds] = - \Lambda^\ast (t).  
\end{equation}

Then, if condition   (\ref{sufficient-condition}) is met, the disentangled structure of the unitary operator (\ref{T-unitary-operator}) is

\begin{equation}
\hat{T} (t,0) =   e^{     \tan [\int_{0}^{t} \|  w_{12} (\tau) \|  d\tau  ] \;  e^{i \phi_0 }   \;  \hat{J}_+}   
e^{     2 \ln \left(  \sec  [\int_{0}^{t} \|  w_{12} (\tau) \|  d\tau  ]   \right)                   \hat{J}_{3} } 
e^{ -  \tan [\int_{0}^{t} \|  w_{12} (\tau) \|  d\tau  ] \;  e^{-i \phi_0 }  \hat{J}_{-} },
\end{equation}
where  we have used  $\| \eta (\tau) \| = \| w_{12} (\tau)\|, \; \forall \; \tau,$  according to  the relation $\eta (\tau) = -i w_{12} (\tau) e^{i \int_{0}^{\tau} (w_{11} (s)  - w_{22} (s) ) ds  }.$ 
It is easy now to rewrite this last operator in the usual form

\begin{equation}
\hat{T} (t) = \exp\left[  \left( \int_{0}^{t} \|  w_{12} (\tau) \|  d\tau \right)  \left( e^{i \phi_0} \hat{J}_+ -  e^{-i \phi_0}  \hat{J}_-  \right)   \right].
\end{equation}

Thus the completely disentangled form of  (\ref{second-Unitary-operator}) writes

\begin{eqnarray}
\hat{U}_{0} (t,0) &=&  \exp\left[{ - i \int_{0}^{t} (w_{11} (s) + w_{22} (s)  ) ds \; \hat{N}}\right]   \; \exp\left[{- i \int_{0}^{t} (w_{11} (s) - w_{22} (s)  ) ds  \; \hat{J}_3} \right] \nonumber \\ &\times& \exp\left[  \left(\int_{0}^{t} \|  w_{12} (\tau) \|  d\tau  \right)  \left( e^{i \phi_0} \hat{J}_+ -  e^{-i \phi_0}  \hat{J}_-  \right)   \right]. \label{U0-untangled}
 \end{eqnarray}

\subsubsection{An explicit untangled form of the time evolution operator}
In our case, when the condition (\ref{sufficient-condition}) is met, which according to the $\eta (s)$ definition and to  ( \ref{condition-phase-constant}) is equivalent to

\begin{equation}
\theta_{12} (t) = - \theta_{21} (t) =  \phi_{0} + \frac{\pi}{ 2} - \int_{0}^{t} (w_{11} (s) - w_{22}(s)) ds, \label{condition-phase-12}
\end{equation}
with no other constraint on the $w_{\sigma \lambda } (t)$ functions, then  the time evolution operator  (\ref{U-starting-non-disentangled}) can be completely  disentangled.  In this case, the solutions of the system (\ref{eq-sys-differential}),  as it has been shown in the above section,  can be obtained by using the matrix elements of the unitary  operator  (\ref{unitary-operator-solution}) in a given orthonormal basis. As 

\begin{equation}
\hat{W} (s) = (w_{11} (s) + w_{22} (s)) \hat{N} + (w_{11} (s) - w_{22} (s)) \hat{J}_3 + w_{12} (s) \hat{J}_+ +  w_{21} (s) \hat{J}_-, 
\end{equation}
a suitable basis to compute the matrix elements of  (\ref{unitary-operator-solution}) is the $j= \frac{1}{2}$ basis representation of the $su(2)$ Lie algebra: $ \mid \pm \rangle = \mid \frac{1}{2} , \pm  \frac{1}{2} \rangle.$ On the other hand, as  the unitary operator  (\ref{unitary-operator-solution}) is identical to the unitary operator (\ref{second-Unitary-operator}), its untangled form is given by equation (\ref{U0-untangled}). Thus, the matrix elements of $\hat{S} (t,0)$ are given by

\begin{align}
S_{11} (t) = \langle + \mid \hat{S} (t,0) \mid + \rangle = e^{- i  \int_{0}^{t}  w_{11} (s) ds  } \cos\left( \int_{0}^{t}  \| w_{12} (s) \| ds       \right),  \\  \quad  S_{12} (t) = \langle + \mid \hat{S} (t,0) \mid - \rangle = e^{- i  \int_{0}^{t}  w_{11} (s) ds  }  \;  e^{i \phi_0}  \; \sin\left( \int_{0}^{t}  \| w_{12} (s) \| ds \right),         \\
\quad  S_{21} (t) = \langle -  \mid \hat{S} (t,0) \mid + \rangle = -  e^{- i  \int_{0}^{t}  w_{22} (s) ds  }  \;  e^{-i \phi_0}  \; \sin\left( \int_{0}^{t}  \| w_{12} (s) \| ds \right), \\
\quad  S_{22} (t) = \langle - \mid \hat{S} (t,0) \mid - \rangle = e^{- i  \int_{0}^{t}  w_{22} (s) ds  } \cos\left( \int_{0}^{t}  \| w_{12} (s) \| ds \right).          
\end{align}

By inserting these last matrix elements in equation  (\ref{c-solutions}), which gives us the  $c_{\sigma (t)}, \; \sigma=1,2,$  functions that  
verify the differential equation system (\ref{eq-sys-differential}), we get

\begin{eqnarray}
c_{1} (t) &= &e^{- i  \int_{0}^{t}  w_{11} (s) ds  } \cos\left( \int_{0}^{t}  \| w_{12}  (s) \| ds       \right) \tilde{c}_{1} (0)  \nonumber \\ &+&
e^{- i  \int_{0}^{t}   w_{11} (s) ds  } \;  e^{i \phi_0} \;  \sin\left( \int_{0}^{t}  \| w_{12} (s) \| ds       \right) \tilde{c}_{2} (0)
\nonumber \\
& -& 
\frac{i}{\hbar}   e^{- i  \int_{0}^{t}  w_{11} (s) ds  }  \int_{0}^{t} e^{i  \int_{0}^{s}  w_{11} (\tau) d\tau  }  \cos\left( \int_{s}^{t}  \| w_{12}  (\tau) \| d\tau       \right) F_{1} (s)  \nonumber \\ & -& \frac{i}{\hbar}   e^{ -i \int_{0}^{t}  w_{11} (s) ds  } \;  e^{i \phi_0}   \;  \int_{0}^{t} e^{i  \int_{0}^{s}  w_{22} (\tau) d\tau  }  \sin\left( \int_{s}^{t}  \| w_{12}  (\tau) \| d\tau \right) F_2 (s) 
\end{eqnarray} 
and

\begin{eqnarray}
c_{2} (t) &= & - \;  e^{- i  \int_{0}^{t}  w_{22} (s) ds  }  \; e^{-i \phi_0}  \;   \sin\left( \int_{0}^{t}  \| w_{12} (s) \| ds       \right) \tilde{c}_{1} (0)  \nonumber \\ &+&
e^{- i  \int_{0}^{t}  w_{22} (s) ds  } \cos\left( \int_{0}^{t}  \| w_{12} (s)  \| ds       \right) \tilde{c}_{2} (0)
\nonumber \\
& + & 
\frac{i}{\hbar}    e^{- i  \int_{0}^{t}  w_{22} (s) ds  }  \;  e^{-i \phi_0}  \;\int_{0}^{t} e^{i  \int_{0}^{s}  w_{11} (\tau) d\tau  }  \sin\left( \int_{s}^{t}  \| w_{12}  (\tau) \| d\tau       \right) F_{1} (s)  \nonumber \\ & -& \frac{i}{\hbar}   e^{- i  \int_{0}^{t}  w_{22} (s) ds }  \int_{0}^{t} e^{i  \int_{0}^{s}  w_{22} (\tau) d\tau  }  \cos\left( \int_{s}^{t}  \| w_{12}  (\tau) \| d\tau \right) F_2 (s). 
\end{eqnarray} 
With these results, we now are sure that a  closed disentangled form of the unitary time evolution operator (\ref{U-starting-non-disentangled}) can be  reached, provided that the phase  condition (\ref{condition-phase-12}) be satisfied. 

\subsection{The case $\eta (s) = \eta_{0} e^{i ( \phi_0 + w_{0} s)}, \;  \eta_{0}, \phi_0  $ and $w_{0}$ constant}
Let us now consider the case when $\eta (s)  = \eta_{0} e^{i (\phi_0 + w_{0}  s)}, $ where  $ \eta_{0} , \phi_0 $ and $w_{0}$ are given constants.
As $\eta (s) = - i w_{12} e^{i \rho (s)},$ this  implies that $\| w_{12} (s) \| =\eta_0$ and
\begin{equation}
\theta_{12} (s) = \phi_0 + \frac{\pi}{2} + w_0 \; s - \int_{0}^{s} (w_{11} (\tau) -w_{22} (\tau) ) d\tau.  \label{phase-12-condition-1}
\end{equation}
Let us now note that here this condition can be interpreted in both directions, that is, for given values of the time-dependent parameters  $w_{11} (s)$ and  $w_{22} (s)$ we must choose $\theta_{12} (s)$  such that the condition is met, and vice versa,   or  force all them to meet the condition and thus obtain fixed values of the constants $\phi_0$ and $w_0.$  So, for example, in the second case, we can choose $\theta_{12} (s) = \nu_{12} \; s, $ and  $w_{11} (s) -w_{22} (s)=\nu_{0}, $ where $\nu_{12}$ and $\nu_{0}$ are constants, and 
obtain the fixed values $\phi_0 = - \frac{\pi}{2} $ and $w_0 = \nu_0 + \nu_{12.}$

Then, taking into account condition (\ref{phase-12-condition-1}) in equation (\ref{eq-Riccati}), we get

\begin{equation}
e^{-i ( \phi_0 + w_{0} s )} \frac{d}{ds} \Lambda (s) = \eta_{0} \left[  1  + \left( e^{-i (\phi_0 + w_0 s)}  \Lambda (s) \right)^2 \right], \quad \Lambda (0)=0. 
\end{equation} 
Now, by performing the  change of variables 
\begin{equation}
\Lambda (s ) =   e^{i(\phi_0 + w_0 s)} \tilde{\Lambda} (s),  \label{variable-change-1}
\end{equation}
this last equation can be put in the form:
\begin{equation}
\frac{2 \eta_0}{\delta} \; \frac{d}{ds} \tilde{\Lambda} (s) =  \frac{ \delta}{2}  \left[  1  +  \left(\frac{ 2 \eta_0}{ \delta}\right)^2  \left( \tilde{\Lambda} (s) -  \frac{ i w_0 }{2 \eta_0} \right)^2 \right] , \quad \tilde{\Lambda} (0) =0, \quad \label{tilde-Lambda-equation}
\end{equation}
where $\delta= \sqrt{4 \eta_0^2 + w_0^2}. $
By performing again a new change of variables:

\begin{equation}
\tilde{\Upsilon} (s) = \frac{ 2 \eta_0 }{\delta}  \left( \tilde{\Lambda} (s) -  \frac{ i w_0 }{2 \eta_0} \right), \label{variable-change-2}
\end{equation}
equation (\ref{tilde-Lambda-equation}) take the form

\begin{equation}
 \frac{d}{ds} \tilde{\Upsilon} (s) = \frac{\delta}{  2} \left[ 1  + \left(\tilde{\Upsilon} (s)\right)^2 \right], \quad \tilde{\Upsilon} (0) = - \frac{i w_0}{\delta},
\end{equation}
which  is easily separable and leads to the solution

\begin{equation} 
\tilde{\Upsilon} (s) = \frac{\delta \tan{\left( \frac{\delta s}{2}\right)}  - i w_0 }{  \delta + i  w_0 \tan \left( \frac{\delta s}{2}\right)  }.
\end{equation}
Returning to the original variable with the help of (\ref{variable-change-1}) and (\ref{variable-change-2}), after some manipulations, we get

\begin{equation}
\Lambda (s) = \frac{2 \eta_0  e^{i (\phi_0  + w_0 s)}  \tan{ \left( \frac{\delta s}{2}\right)}  }{\sqrt{ \delta^2 + w_0^2  \tan^2{ \left( \frac{\delta s}{2}\right)}   } } \exp\left[ - i     \arctan \left[ \frac{w_0}{ \delta} \tan\left( \frac{\delta \; s}{2} \right)  \right]        \right]
\end{equation}
Finally, by inserting this last result into (\ref{Omega-Gamma-expressions}), then integrating and   manipulating the resulting  expressions we obtain
 \begin{equation}
 \Omega (s) = \ln\left[ \frac{\delta^2 \sec^2 \left( \frac{\delta \, s }{2} \right)}{ \delta^2 + w_0^2  \tan^2 \left( \frac{\delta \; s}{2}\right)   }  \right]  + i \left[ w_0 \;  s - 2  \arctan \left[ \frac{w_0}{ \delta} \tan\left( \frac{\delta \; s}{2} \right)  \right]  \right]   
 \end{equation}
and
\begin{equation}
\Gamma (s) = - \frac{2 \eta_0 e^{- i \phi_0}   \tan\left( \frac{\delta \; s}{2} \right)}{ \sqrt{ \delta^2 + w_0^2  \tan^2 \left( \frac{\delta \; s}{2}\right)   } } \exp\left[ - i     \arctan \left[ \frac{w_0}{ \delta} \tan\left( \frac{\delta \; s}{2} \right)  \right]        \right].
\end{equation}

Finally, by inserting these last results into  (\ref{T-su2-dis-3}), and using the general relation $e^{\alpha \hat{J}_3}  \hat{J}_{-} e^{- \alpha \hat{J}_3} = e^{- \alpha}  \hat{J}_{-}, $ we obtain the untangle form of the $\hat{T} (t)$ unitary operator, that is

\begin{eqnarray}
\hat{T} (t) &=&  \exp \left[ \frac{2 \eta_0  e^{i (\phi_0  + w_0 t)}  \tan{ \left( \frac{\delta t}{2}\right)}  }{\sqrt{ \delta^2 + w_0^2  \tan^2{ \left( \frac{\delta t}{2}\right)}   } } \exp\left[ - i     \arctan \left[ \frac{w_0}{ \delta} \tan\left( \frac{\delta \; t}{2} \right)  \right]        \right] \;  \hat{J}_{+} \right]  \nonumber \\ &\times& 
\exp \left[        \ln\left[ \frac{\delta^2 \sec^2 \left( \frac{\delta \, t }{2} \right)}{ \delta^2 + w_0^2  \tan^2 \left( \frac{\delta \; t}{2}\right)   }  \right]                     \; \hat{J}_{3}  \right]    \exp \left[ \frac{- 2 \eta_0  e^{- i (\phi_0  + w_0 t)}  \tan{ \left( \frac{\delta t}{2}\right)}  }{\sqrt{ \delta^2 + w_0^2  \tan{ \left( \frac{\delta t}{2}\right)}   } }  \exp\left[  i     \arctan \left[ \frac{w_0}{ \delta} \tan\left( \frac{\delta \; t}{2} \right)  \right]        \right] \;  \hat{J}_{-} \right] \nonumber \\ &\times&\exp\left[   i \left[ w_0 \;  t - 2  \arctan \left[ \frac{w_0}{ \delta} \tan\left( \frac{\delta \; t}{2} \right)  \right]  \right]      \; \hat{J}_{3}  \right], 
\end{eqnarray} 
which written in the standard form look like

\begin{eqnarray}
\hat{T} (t) &=&  \exp \left\{   \arctan\left( 
\frac{2 \eta_0  \tan{ \left( \frac{\delta t}{2}\right)}  }{\sqrt{ \delta^2 + w_0^2  \tan^2 { \left( \frac{\delta t}{2}\right)}   } }
\right)   \right. \nonumber  \\  & \times &
\left.
\left[    e^{ i \left( \phi_0  + w_0 \;  t   - \arctan \left( \frac{w_0}{ \delta} \tan\left( \frac{\delta \; t}{2} \right) \right)  \right) }\; \hat{J}_{+}  -     e^{ - i \left( \phi_0  + w_0 \;  t   - \arctan \left( \frac{w_0}{ \delta} \tan\left( \frac{\delta \; t}{2} \right) \right) \right) }\; \hat{J}_{-}      \right] \right\}  \nonumber \\
 &\times& e^{  i \left[ w_0 \;  t - 2  \arctan \left( \frac{w_0}{ \delta} \tan\left( \frac{\delta \; t}{2} \right)  \right)  \right]      \; \hat{J}_{3} }. \label{T-standard-1}
\end{eqnarray}

As before, so that untangled form of the time evolution is reached, we need  to compute the $S$ matrix elements of unitary operator (\ref{second-Unitary-operator}). For this we can use again the vector basis of the  $j=\frac{1}{2}$ irreducible representation of su(2) algebra, doing that we obtain

\begin{eqnarray}
S_{11} (t) &=& \langle + \mid \hat{S} (t,0) \mid + \rangle =   e^{- i  \int_{0}^{t}  w_{11} (s) ds  }  
 \; e^{ i \frac{w_0 \; t }{2} }        \left[  \cos\left( \frac{\delta t}{2} \right) 
 - i \frac{w_0}{\delta}   \sin\left( \frac{\delta \; t}{2} \right) \right] 
 \nonumber
 \\ S_{12} (t)  &=& \langle + \mid \hat{S} (t,0) \mid - \rangle =  \frac{ 2 \eta_0}{ \delta} e^{- i  \int_{0}^{t}  w_{11} (s) ds  }  \;  e^{i (\phi_0 + \frac{w_0 \; t }{2})}  \; \sin\left( \frac{\delta \; t}{2} \right),        \nonumber \\   S_{21} (t) &=& \langle - \mid \hat{S} (t,0) \mid + \rangle = -  \frac{ 2 \eta_0}{ \delta} e^{- i  \int_{0}^{t}  w_{22} (s) ds  }  \;  e^{-i (\phi_0 + \frac{w_0 \; t }{2})}  \; \sin\left( \frac{\delta \; t}{2} \right) ,  \nonumber\\ S_{22} (t) &=& \langle + \mid \hat{S} (t,0) \mid + \rangle =
     e^{- i  \int_{0}^{t}  w_{22} (s) ds  } 
 \; e^{ - i \frac{w_0 \; t }{2} }        \left[  \cos\left( \frac{\delta t}{2} \right) 
 +  i \frac{w_0}{\delta}   \sin\left( \frac{\delta \; t}{2} \right) \right] . \nonumber \\
\end{eqnarray}

\subsection{The case $\eta (s) =  r(s) e^{i \phi(s)} $ with $ r (s)=  \epsilon \frac{\eta_0 }{w_0}   \frac{d\phi}{ds} (s)   >  0 \; \forall s,   $ where    $\epsilon = \pm 1  $ and $\eta_0$ and $w_0 $ are   constants}
The last  case treated above can be generalized if  the norm of $\eta(s)$ and its phase verify the following condition
 
  \begin{equation}
r (s) =   \epsilon \; \frac{\eta_0 }{w_0}  \frac{d\phi}{ds} (s) =   \begin{cases}
+   \eta_0  \frac{d\phi (s)}{ds} ,  & \text{if $\frac{d\phi (s)}{ds} > 0$}.\\
 -  \frac{\eta_0 }{w_0}  \frac{d\phi (s)}{ds} , &  \text{if $\frac{d\phi (s)}{ds} < 0$} .
  \end{cases}
\end{equation}
where $\eta_0$ and $w_0$ are given non-negative real constants greater than zero which have been written here in this way  in order to  make  compatible the physical quantities.  According the  definition of the parameter $\eta (s),$ this condition is equivalent to

\begin{equation}
\theta_{12} (t) = - \theta_{21} (t) =  \phi (t )+ \frac{\pi}{ 2} - \int_{0}^{t} (w_{11} (s) - w_{22}(s)) ds, \label{new-condition-phase-12}
\end{equation}
 and
\begin{equation}
\| w_{12} (s) \| = \epsilon  \frac{\eta_0}{w_0}  \frac{d\phi}{ds} (s).    
\end{equation} 
Assuming that holds,  then  (\ref{eq-Riccati}) can be analytically  solved. Indeed, by  performing the change of variables:

\begin{equation}
\tilde{\Upsilon} (s) = \frac{ 2 \eta_0 }{\delta}  \left( e^{- i \phi(s)} \; {\Lambda} (s) - i   \frac{w_0 }{2 \epsilon \; \eta_0} \right), \label{variable-change-three}
\end{equation}
where $\delta=\sqrt{4 \eta_0^2 +w_0^2}, $ equation (\ref{eq-Riccati}) take the form

\begin{equation}
 \frac{d}{ds} \tilde{\Upsilon} (s) = \frac{\delta \; \epsilon}{2 w_0}  \frac{d\phi}{ds} (s)    \left[ 1  + \left(\tilde{\Upsilon} (s)\right)^2 \right], \quad \tilde{\Upsilon} (0) = -i \frac{ w_0}{\delta \; \epsilon},
\end{equation}
which  is easily separable and leads to the solution

\begin{equation} 
\tilde{\Upsilon} (s) =  \epsilon  \; \frac{\delta \tan{\left(    \frac{\delta \; \tilde{\phi} (s)}{2 w_0}  \right)}  -  i w_0 }{  \delta +  i  w_0 \tan \left(   \frac{\delta \; \tilde{\phi} (s)}{2 w_0}    \right)  }, \label{solution-Upsilon-abs-values}
\end{equation}
 where $\tilde{\phi} (s) =  \phi (s) - \phi(0) $ and the $\epsilon$ parameter value stands for the region  where $  \epsilon \frac{d\phi}{ds} (s)  > 0.$  

Returning to the original variable with the aid of  (\ref{variable-change-three}) and (\ref{solution-Upsilon-abs-values}), after some manipulations, we get

\begin{equation}
\Lambda (s) =  \frac{2 \; \epsilon \eta_0  e^{i  \phi( s)}  \tan{ \left( \frac{\delta \; \tilde{\phi} (s)}{2 w_0}    \right)}  }{\sqrt{ \delta^2 + w_0^2  \tan^2{ \left( \frac{\delta \;\tilde{\phi (s)}  }{2 w_0}  \right)}   } } \exp\left[ - i     \arctan \left[ \frac{w_0}{ \delta} \tan\left( \frac{\delta \; \tilde{\phi} (s) }{2w_0}  \right)  \right]        \right],
\end{equation}
 
Thus, by inserting this last result into (\ref{Omega-Gamma-expressions}) and then integrating and   manipulating the resulting  expressions we obtain
 \begin{equation}
 \Omega (s) = \ln\left[ \frac{\delta^2 \sec^2 \left( \frac{\delta \; \tilde{\phi}(s)}{2 w_0}    \right)}{ \delta^2 + w_0^2  \tan^2 \left(\frac{\delta \;  \tilde{\phi}(s)}{2 w_0}    \right)   }  \right]  + i \left[ \tilde{\phi} (s) - 2  \arctan \left[ \frac{w_0}{ \delta} \tan\left(   \frac{\delta \; \tilde{\phi}(s)}{2 w_0}             \right)  \right]  \right]   
 \end{equation}
and
\begin{equation}
\Gamma (s) = -  \frac{2 \; \epsilon  \eta_0 e^{- i  \phi (0) }   \tan\left(   \frac{\delta \; \tilde{\phi}(s)}{2 w_0}     \right)}{ \sqrt{ \delta^2 + w_0^2  \tan^2 \left( \frac{\delta \; \tilde{\phi}(s)}{2 w_0}   \right)   } } \exp\left[ - i     \arctan \left[ \frac{w_0}{ \delta} \tan\left(   \frac{\delta \; \tilde{\phi}(s)}{2 w_0}     \right)  \right]        \right].
\end{equation}

Finally, with the help of  all these expression we can  write the unitary operator $\hat{T} (t)$ in the form
 
   \begin{eqnarray}
\hat{T} (t) &=& \exp \left[ \frac{  2  \; \epsilon \; \eta_0  e^{i  \phi (t) }  \tan{ \left( \frac{\delta \;  \tilde{\phi}(t)}{2 w_0}  \right)}  }{\sqrt{ \delta^2 + w_0^2  \tan^2 { \left( \frac{\delta \;  \tilde{\phi}(t)}{2 w_0}  \right)}   } } \exp\left[ - i     \arctan \left[ \frac{w_0}{ \delta} \tan\left( \frac{\delta \;  \tilde{\phi}(t)}{2 w_0}   \right)  \right]        \right] \;  \hat{J}_{+} \right]  \nonumber \\ &\times& 
\exp \left[        \ln\left[ \frac{\delta^2 \sec^2 \left(\frac{\delta \;  \tilde{\phi}(t)}{2 w_0}   \right)}{ \delta^2 + w_0^2  \tan^2 \left(  \frac{\delta \;  \tilde{\phi}(t)}{2 w_0}  \right)   }  \right]                     \; \hat{J}_{3}  \right]  \nonumber \\ &\times&  \exp \left[ \frac{ -   \;  2 \; \epsilon \; \eta_0  e^{- i  \phi  (t) }  \tan{ \left( \frac{\delta \;  \tilde{\phi}(t)}{2 w_0}  \right)}  }{\sqrt{ \delta^2 + w_0^2  \tan^2 { \left( \frac{\delta \;  \tilde{\phi}(t)}{2 w_0}  \right)}   } }  \exp\left[  i     \arctan \left[ \frac{w_0}{ \delta} \tan\left( \frac{\delta \;  \tilde{\phi}(t)}{2 w_0}   \right)  \right]        \right] \;  \hat{J}_{-} \right] \nonumber \\ &\times&\exp\left[   i \left[  \tilde{\phi} (t) - 2  \arctan \left[ \frac{w_0}{ \delta} \tan\left(  \frac{\delta \;  \tilde{\phi}(t)}{2 w_0}    \right)  \right]  \right]      \; \hat{J}_{3}  \right], 
\end{eqnarray} 
or in the the standard form, as we have done in the above sections:

\begin{eqnarray}
\hat{T} (t) &=&  \exp \left\{   \arctan\left( 
\frac{   2  \; \epsilon \; \eta_0  \tan{ \left(\frac{\delta \;  \tilde{\phi}(t)}{2 w_0}   \right)}  }{\sqrt{ \delta^2 + w_0^2  \tan^2{ \left(  \frac{\delta \;  \tilde{\phi}(t)}{2 w_0}      \right)}   } }
\right)   \right. \nonumber  \\  & \times &
\left.
\left[    e^{ i \left(   \phi(t)  - \arctan \left( \frac{w_0}{ \delta} \tan\left(\frac{\delta \;  \tilde{\phi}(t)}{2 w_0}    \right) \right)  \right) }\; \hat{J}_{+}  -     e^{ - i \left(    \phi (t)  - \arctan \left( \frac{w_0}{ \delta} \tan\left(  \frac{\delta \;  \tilde{\phi}(t)}{2 w_0}     \right) \right) \right) }\; \hat{J}_{-}      \right] \right\}  \nonumber \\
 &\times& e^{  i \left[ \tilde{\phi} (t) - 2  \arctan \left( \frac{w_0}{ \delta} \tan\left( \frac{\delta \;  \tilde{\phi}(t)}{2 w_0}   \right)  \right)  \right]      \; \hat{J}_{3} }, \label{T-standard-2}
\end{eqnarray}
which reduces to (\ref {T-standard-1}) when $\phi (s) =  \phi_0 + w_0 \; s. $

As before we can compute  the $S$ matrix elements of  (\ref{second-Unitary-operator}) in the $j=\frac{1}{2}$  representation space of the $su(2)$ Lie algebra. They are given by

\begin{eqnarray}
S_{11} (t) & =&   e^{- i  \int_{0}^{t}  w_{11} (s) ds  } \;  
e^{ i  \frac{\tilde{\phi} (t)}{2} }  \;  
 \left[ \cos\left( \frac{\delta \;  \tilde{\phi}(t)}{2 w_0}  \right) 
-i \frac{w_0}{\delta} \; \sin\left( \frac{\delta \;  \tilde{\phi}(t)}{2 w_0}  \right) \right] 
\nonumber
 \\ S_{12} (t)  & =&   \;  \frac{  2 \;  \epsilon \; \eta_0}{ \delta} e^{- i  \int_{0}^{t}  w_{11} (s) ds  }  \;  e^{i  \frac{(\phi (t) + \phi(0))}{2}  }  \; \sin\left( \frac{\delta \;  \tilde{\phi}(t)}{2 w_0}     \right),    \nonumber \\   S_{21} (t) &=&  -  \;   \frac{ 2 \; \epsilon \;  \eta_0}{ \delta} e^{- i  \int_{0}^{t}  w_{22} (s) ds  }  \;  e^{-i \frac{(\phi (t) + \phi (0)) }{2} }  \; \sin\left(   \frac{\delta \;  \tilde{\phi}(t)}{2 w_0}   \right) ,  \nonumber\\
 S_{22} (t) & =&      e^{- i  \int_{0}^{t}  w_{22} (s) ds  } \; e^{ - i  \frac{\tilde{\phi} (t)}{2} }  \;  
 \left[ \cos\left( \frac{\delta \;  \tilde{\phi}(t)}{2 w_0}  \right) 
+ i \frac{w_0}{\delta} \; \sin\left( \frac{\delta \;  \tilde{\phi}(t)}{2 w_0}  \right) \right], \nonumber \\ \label{S-matrix-elements-phi-derivative}
\end{eqnarray} 
where in all the above expressions the sing of $\epsilon$ accords with the sign of $\frac{d \phi}{ds} (s).$

\subsection{The standard case when all parameters $w_{\sigma \lambda} $  are  time-independent} 
When all parameters $w_{\sigma \lambda} $ are time-independent, i.e., $w_{11} (s) = w_{11},$ $w_{22} (s) = w_{22},$ and $w_{12} (s)= w_{21}^\ast (s)=  \|w_{12} \| \; e^{i \theta^0_{12}},$  where $w_{11} \neq w_{22}$     and $\theta^0_{12} \in [0,2 \pi],$  
the condition (\ref{new-condition-phase-12}) implies

\begin{equation}
\theta^0_{12}  = - \theta^0_{21}  =  \phi (s )+ \frac{\pi}{ 2} -  (w_{11} -w_{22} ) s, \label{condition-all-constants}
\end{equation}
 from were 
\begin{equation}
\phi (s) = \theta^0_{12} - \frac{ \pi}{2} + (w_{1	1} - w_{22}) s, \quad {\rm and } \quad  \tilde{\phi} (s) = (w_{11} - w_{22}) s. 
\end{equation} 			
 
 Thus the different combinations of parameters are 

\begin{equation}	
 \frac{\eta_0 }{w_0}	 = \frac{\epsilon \; \|  w_{12} \|} {(w_{11} -w_{22})}, \quad    
\delta = \frac{\epsilon \; w_{0} \; b}{ (w_{11} - w_{22})} \quad {\rm and} \quad   \frac{\delta \tilde{\phi } (s)}{2 w_0}   = \frac{ \epsilon \; b \; s}{2},
\end{equation} 
where $b= \sqrt{ 4 \| w_{12} \|^2 + (w_{11} - w_{22})^2}.$ Hence, the  unitary operator $\hat{T} (t)$ takes the form

 \begin{eqnarray}
\hat{T} (t) &=&  \exp \left[ \frac{  2    w_{12}   e^{i (w_{11} -w_{22}) t - i \frac{\pi}{2} }  \tan{ \left( \frac{b \; t}{2}            \right)}  } {\sqrt{  b^2+ (w_{11} - w_{22})^2   \tan^2 { \left(   \frac{b \; t}{2}    \right)}   } }\exp\left[ - i     \arctan \left[  \frac{(w_{11} -w_{22})}{ b} \tan\left( \frac{b \; t}{2}  \right)  \right]        \right] \;  \hat{J}_{+} \right] \nonumber \\ &\times& 
\exp \left[        \ln\left[ \frac{b^2 \sec^2 \left( \frac{b \; t}{2} \right)}{ b^2 + (w_{11}  - w_{22})^2 \tan^2 \left(  \frac{b \; t}{2}  \right)   }  \right]                     \; \hat{J}_{3}  \right]  \nonumber \\ &\times&  \exp \left[ \frac{ -  2    w_{21}   e^{- i (w_{11} -w_{22}) t + i \frac{\pi}{2} }  \tan{ \left( \frac{b \; t}{2}            \right)}  } {\sqrt{  b^2+ (w_{11} - w_{22})^2   \tan^2 { \left(   \frac{b \; t}{2}    \right)}   } }\exp\left[  i     \arctan \left[  \frac{(w_{11} - w_{22} )}{ b} \tan\left( \frac{b \; t}{2}  \right)  \right]        \right] \;  \hat{J}_{-} \right] \nonumber \\ &\times&\exp\left[   i \left[  (w_{11} -w_{22})  (t) -  2  \arctan \left[ \frac{(w_{11} - w_{22})}{ b} \tan\left(  \frac{ b \; t}{2}   \right)  \right]  \right]      \; \hat{J}_{3}  \right], 
\end{eqnarray} 

Thus the operator (\ref{U0-untangled}) becomes

\begin{eqnarray}
\hat{U}_{0} (t,0) &=& \exp\left[ - i (w_{11} + w_{22}) \hat{N}  \right] \nonumber \\ &\times&
\exp \left[ \frac{ 2  w_{12}   \tan{ \left( \frac{b \; t}{2}            \right)}  } {\sqrt{  b^2+ (w_{11} - w_{22})^2   \tan^2 { \left(   \frac{b \; t}{2}    \right)}   } }\exp\left[ - i \frac{\pi}{2}  - i     \arctan \left[ \arctan\left(  \frac{(w_{11} -w_{22})}{ b} \tan\left( \frac{b \; t}{2}  \right) \right)  \right]        \right] \;  \hat{J}_{+} \right] \nonumber \\ &\times& 
\exp \left[        \ln\left[ \frac{b^2 \sec^2 \left( \frac{b \; t}{2} \right)}{ b^2 + (w_{11}  - w_{22})^2 \tan^2 \left(  \frac{b \; t}{2}  \right)   }  \right]                     \; \hat{J}_{3}  \right]  \nonumber \\ &\times&  \exp \left[ \frac{ -  2    w_{21}  \tan{ \left( \frac{b \; t}{2}            \right)}  } {\sqrt{  b^2+ (w_{11} - w_{22})^2   \tan^2 { \left(   \frac{b \; t}{2}    \right)}   } }\exp\left[ i  \frac{\pi}{2}  + i     \arctan \left[  \frac{(w_{11} - w_{22} )}{ b} \tan\left( \frac{b \; t}{2}  \right)  \right]        \right] \;  \hat{J}_{-} \right] \nonumber \\ &\times&\exp\left[  -  2 i   \arctan \left[ \frac{(w_{11} - w_{22})}{ b} \tan\left(  \frac{ b \; t}{2}   \right)  \right]       \; \hat{J}_{3}  \right],
\end{eqnarray} 
where we can recognize the standard  Bigolovuv  disentangled form of an $su(2)$ exponential operator, and the $\hat{S} (t,0)$ matrix elements take the form

\begin{eqnarray}
S_{11} (t) & =&   e^{- i \frac{ (w_{11} + w_{22}) \; t }{2}    } \;  
 \left[ \cos\left( \frac{b t}{2} \right) 
-i \frac{(w_{11} - w_{22})}{b} \; \sin\left( \frac{b t}{2}  \right) \right] 
\nonumber
 \\ S_{12} (t)  & =&   -   \frac{  2 i w_{12}}{ b}   e^{- i  \frac{(w_{11} + w_{22}) \; t}{2}   } \; \sin\left( \frac{b t}{2}  \right),  
  \nonumber \\   S_{21} (t) &=&   -   \frac{  2 i w_{21}}{ b}   e^{- i \frac{  (w_{11} + w_{22}) \; t }{2}  } \; \sin\left( \frac{b t}{2}  \right),  
 \nonumber\\
 S_{22} (t) & =&   e^{- i \frac{ (w_{11} + w_{22}) \; t}{2}   } \;  
 \left[ \cos\left( \frac{b t}{2} \right) 
+ i \frac{(w_{11} - w_{22})}{b} \; \sin\left( \frac{b t}{2}  \right) \right].   \label{S-matrix-all-constants}
\end{eqnarray}
 
\section{Time evolution of the coherent states linked to  the   commensurate 2 D isotropic harmonic oscillator}
\label{sec-four}

The Schr\"odinger type coherent states  of the  two-dimensional isotropic harmonic oscillator 
\begin{equation} 
H_0 = \hbar \hat{a_1}^\dagger \hat{a}_1 + \hbar  \hat{a_2}^\dagger \hat{a}_2 \end{equation} \label{H-0-1-1}
can be defined as the  eigentates of the two-mode annihilation operator
\begin{equation}
\hat{A}_{\alpha \beta} = \alpha \hat{a}_1  + \beta \hat{a}_2, \label{A-alpha-beta-constants}
\end{equation}
where $ \alpha$ and $\beta $ are complex numbers which verify $\|\alpha\|^2 + \|\beta\|^2 =1.$  
Indeed, due to this last fact, the operators $\hat{H}_0,$   $\hat{A}_{\alpha \beta}$ and   its adjoint $\hat{A}^\dagger_{\alpha \beta}$ and the identity $\hat{I}$ satisfy  commutation relations of the oscillator algebra 

\begin{equation} \left[ \hat{A}_{\alpha \beta} ,
\hat{A}^\dagger_{\alpha \beta} \right]=\hat{I},
\end{equation} 
and 
\begin{equation}
[\hat{H}_{0} ,   \hat{A}_{\alpha \beta}] = - \hbar \; \hat{A}_{\alpha \beta} \quad {\rm and} \quad [\hat{H}_{0} ,   \hat{A}^\dagger_{\alpha \beta}] = \hbar  \hat{A}^\dagger_{\alpha \beta}. 
\end{equation}

The Hamiltonian (\ref{H-0-1-1}) is not the only  $h(1) \oplus h(1) \roplus  u(2)$-type Hamiltonian that possess  these characteristics \cite{NEAM-june-2023}, but also, for example, the following Hamiltonian formed from the product of the ladder operators\cite{NAM-VH}    $\hat{A}^\dagger_{\alpha\beta}$ and $  \hat{A}_{\alpha \beta}:$   

    \begin{equation}
\hat{H}_{\alpha \beta} =  \hbar   \hat{A}^\dagger_{\alpha\beta}  \hat{A}_{\alpha \beta} = \hbar  \; [ \|\alpha \|^2 \hat{a}^\dagger_1\hat{a}_{1}  +
\|\beta \|^2  \hat{a}^\dagger_2 \hat{a}_{2} + \alpha^\ast \beta \hat{a}^\dagger_1 \hat{a}_2  + \alpha \beta^\ast  \hat{a}^\dagger_2 \hat{a}_1 ].  \label{two-mode-attached-hamiltonian}
\end{equation}
Certainly, this Hamiltonian also verify

\begin{equation}
[\hat{H}_{\alpha \beta} ,   \hat{A}_{\alpha \beta}] = - \hbar \; \hat{A}_{\alpha \beta} \quad {\rm and} \quad [\hat{H}_{\alpha \beta} ,   \hat{A}^\dagger_{\alpha \beta}] = \hbar  \hat{A}^\dagger_{\alpha \beta}.  \label{H-alpha-beta-commutation}
\end{equation}

Since the set of operators $ \left\{   \hat{A}_{\alpha \beta}, \hat{A}^\dagger_{\alpha \beta}, \hat{I},  \hat{H}_{\alpha \beta} \right\} $ generate the one dimensional oscillator algebra, we can associate to this system  the one parameter Schr\"odinger type  coherent states\footnote{Indeed, the coherent states in (\ref{coherent-states-Z-2-D}) represent only a special case of a family of algebra eigenstates of the operator  (\ref{A-alpha-beta-constants}) whose coefficients verify the  relation $\|\alpha\|^2 + \|\beta\|^2 =1.$ In this special case, the fundamental state $\mid 0,0 \rangle,$ coincides with the fundamental state of the Hamiltonian   (\ref{two-mode-attached-hamiltonian}) \cite{NEAM-june-2023} .}

\begin{equation}
\mid Z  \rangle  = \exp\left[Z \hat{A}^\dagger_{\alpha \beta} -  Z^\ast  \hat{A}_{\alpha \beta} \right] \mid 0,0 \rangle, \label{coherent-states-Z-2-D}
\end{equation}
 where $\mid 0,0 \rangle$ denotes the two mode quantum oscillator fundamental state, which implies that \linebreak $ \hat{A}_{\alpha \beta} \mid 0,0 \rangle =0.$ 

\vspace{1.0cm}

The time evolution of a state  $\mid \psi (0) \rangle$  governed by the Hamiltonian   (\ref{two-mode-attached-hamiltonian} ) is given by 
\begin{equation}
\mid \psi (t) \rangle = e^{- \frac{i}{\hbar} \int_{0}^{t}  \hat{H}_{\alpha \beta}  ds  } \mid \psi (0) \rangle =   e^{- \frac{i}{\hbar} \hat{H}_{\alpha \beta}  t } \mid \psi (0) \rangle,
\end{equation}
because $ \hat{H}_{\alpha \beta}   $ is time-independent. Thus, the time evolution of the coherent states  \ref{coherent-states-Z-2-D}
is given by
\begin{equation}
\mid Z (t) \rangle =   e^{- \frac{i}{\hbar} \hat{H}_{\alpha \beta}  t }  \mid  Z \rangle=  \exp\left[ -  i  \hat{A}^\dagger_{\alpha\beta}  \hat{A}_{\alpha \beta} \; t \right]   \exp\left[Z \hat{A}^\dagger_{\alpha \beta} -  Z^\ast  \hat{A}_{\alpha \beta} \right] \mid 0,0 \rangle.
\end{equation}
Then, using the generic relations $e^{- i \lambda \hat{a}^\dagger \hat{a}}  \hat{a}^\dagger e^{ i \lambda \hat{a}^\dagger \hat{a}} =  e^{- i \lambda} \hat{a}^\dagger  $  and  $e^{- i \lambda \hat{a}^\dagger \hat{a}}  \hat{a}  e^{ i \lambda \hat{a}^\dagger \hat{a}} =  e^{ i \lambda} \hat{a},$ we get the time evolution 2  D  Shr\"odinger type  coherent states
\begin{equation}
\mid Z (t) \rangle =    \exp\left[ Z e^{- i t} \hat{A}^\dagger_{\alpha \beta} -  Z^\ast e^{i t}  \hat{A}_{\alpha \beta} \right] \mid 0,0 \rangle.
\label{A-alpha-beta-t-coherent-states}
\end{equation}
The same result can be reached if we use the general technique developed in the previous sections. In fact, it has been well established \cite{Su-Me-1967} that the most general time-dependent Hamiltonian that preserves in time the coherent state character of a system
which has initially been prepared as a coherent state  is precisely of the type (\ref{hamiltonian-two-modes}), and that for $N$ interacting quantum harmonic oscillators. Also,  it has been proven \cite{Su-Me-1967}  that the general structure of these coherent states, in the schr\"odinger picture,   is given by

\begin{equation}
\exp\left[\sum_{\sigma =0}^{N} \left(c_{\sigma} (t) \hat{a}^\dagger_\sigma  -  c^\ast_{\sigma} (t) \hat{a}_\sigma \right) \right] \mid \tilde{0} \rangle,  \label{general-coherent-states}
\end{equation}
where the $c_{\sigma} (t)$ coefficients are given by (\ref{c-solutions}) (where the sum is extended to $N$) and   $ \mid \tilde{0} \rangle$ represents the fundamental state of the system.

For the case we are trying now ($N=2, \;  F_{\sigma} (t)= B (t) =0$), the   parameters can be chosen in the following way
\begin{equation}
w_{11} = \|  \alpha \|^2, \quad w_{22} =   \|  \beta \|^2 \quad {\rm and } \quad w_{12}= w^\ast_{21} =  \alpha^\ast \beta,
\end{equation}
from where $w_{11} + w_{22}= 1$ and  $b=1.$  Thus from (\ref{S-matrix-all-constants}) we get the $\hat{S} (t,0)$ matrix elements

\begin{eqnarray}
S_{11} (t) & =&   e^{- i \frac{ t }{2}    } \;  
 \left[ \cos\left( \frac{t}{2} \right) 
-i  \left(\|\alpha\|^2 - \|\beta\|^2 \right) \; \sin\left( \frac{ t}{2}  \right) \right] 
\nonumber
 \\ S_{12} (t)  & =&   -    2  i   \alpha^\ast \beta     e^{- i \frac{ t }{2}    }  \; \sin\left( \frac{b t}{2}  \right),  
  \nonumber \\   S_{21} (t) &=&   -  2 i     \alpha  \beta^\ast  e^{- i \frac{ t }{2}    }  \; \sin\left( \frac{b t}{2}  \right),  
 \nonumber\\
 S_{22} (t) & =&  e^{- i \frac{ t }{2}    }  \;   \left[ \cos\left( \frac{t}{2} \right) 
+ i  \left(\|\alpha\|^2 - \|\beta\|^2 \right) \; \sin\left( \frac{ t}{2}  \right) \right]   
\end{eqnarray}
 and also from  there we obtain the following  $c_{\sigma} (t), \; \sigma=1,2,$ coefficients:
 
 \begin{equation}
 c_{1} (t ) =     e^{- i \frac{ t }{2}    } \;  
 \left[ \cos\left( \frac{t}{2} \right) 
-i  \left(\|\alpha\|^2 - \|\beta\|^2 \right) \; \sin\left( \frac{ t}{2}  \right) \right]  \tilde{c}_{1} (0)  -    2  i   \alpha^\ast \beta     e^{- i \frac{ t }{2}    }  \; \sin\left( \frac{b t}{2}  \right) \tilde{c}_{2} (0) \label{c1-t-2D}
 \end{equation}
 and
  \begin{equation}
   c_{2} (t ) =   -  2 i     \alpha  \beta^\ast  e^{- i \frac{ t }{2}    }  \; \sin\left( \frac{b t}{2}  \right) \tilde{c}_{1} (0)   + e^{- i \frac{ t }{2}    }  \;   \left[ \cos\left( \frac{t}{2} \right) 
+ i  \left(\|\alpha\|^2 - \|\beta\|^2 \right) \; \sin\left( \frac{ t}{2}  \right) \right]  \tilde{c}_{2} (0),\label{c2-t-2D}
 \end{equation}
 which according to the initial condition (\ref{coherent-states-Z-2-D}) verify   $c_{1} (0) = Z \alpha^\ast$ and $c_{2} (0)= Z \beta^\ast.$ This last condition, in this case,  is equivalent to $\tilde{c}_{1} (0) = Z \alpha^\ast$ and $\tilde{c}_{2} (0)= Z \beta^\ast.$ By  inserting these last constants into (\ref{c1-t-2D}) and (\ref{c2-t-2D}) and  simplifying the resulting expressions we get

\begin{equation}
c_{1} (t) = Z \alpha^\ast e^{-i t} \quad {\rm and } \quad c_{2} (t) = Z \beta^\ast e^{-i t}.  \label{c1-c2-t-2-D}
\end{equation} 
  Finally, by inserting (\ref{c1-c2-t-2-D}) into (\ref{general-coherent-states})  (for the particular case $N=2$) and  regrouping suitably the terms, we regain (\ref{A-alpha-beta-t-coherent-states}). 

 \vspace{1.0cm}
 Let us here make some comment on the energy eigenstates and energy spectrum of the Hamiltonian (\ref{two-mode-attached-hamiltonian}). A chain of energy eigenstates for this Hamiltonian can be constructed in the usual way, i.e.,

\begin{equation} 
\mid E_n \rangle = \mathcal{N}_{n; \alpha \beta}^{- \frac{1}{2}}  \left(  \hat{A}^\dagger_{\alpha \beta}  \right)^n \mid 0,0 \rangle,
\end{equation}
 where $ \mathcal{N}_{n; \alpha \beta} = \langle E_n \mid E_n \rangle,$ is a normalization constant. As $\hat{H}_{\alpha \beta} \mid 0,0 \rangle=0,$ then  according to the commutation relations   (\ref{H-alpha-beta-commutation}) the energy spectrum of this system is given by $E_n =n, \; n=0,1,2, \ldots,$ i.e., this system is isospectral to the canonical quantum oscillator. 
 
 \vspace{1.0cm}
 On the other hand, it is interesting to remark that, by means of a unitary similarity transformation,  the  Hamiltonian (\ref{two-mode-attached-hamiltonian}) can be brought to  a simpler form. Indeed, if we write $\hat{H}{\alpha \beta}$ in the standard form

\begin{equation}
\hat{H}_{\alpha \beta} = \hbar [\gamma_{0} \hat{N}   + \gamma_3 \hat{J}_3 +  \gamma_-  \hat{J}_+  +  \gamma_+ \hat{J}_+ ],
\end{equation} 
 where $\gamma_0 = \|\alpha\|^2 + \|\beta\|^2 =1,$ $ \gamma_3 =\|\alpha\|^2 - \|\beta\|^2,$ $\gamma_- = \alpha^\ast \beta$ and
 $\gamma_+ = \alpha \beta^\ast,$ and define $\alpha = \|\alpha\| e^{i \theta_\alpha}$ and $\beta = \| \beta\| e^{i \theta_\beta},$
  the  $su(2)$ unitary mixing operator
\begin{equation}
\hat{T}_{\epsilon} = \exp\left[ - \arctan \left(  \epsilon \sqrt{\frac{1 -\epsilon \gamma_3}{1 + \epsilon \gamma_3}} \right)  
\left( e^{-i (\theta_\alpha - \theta_\beta)} \hat{J}_+ - e^{i (\theta_\alpha - \theta_\beta)}  \hat{J}_- \right)
\right], \quad \epsilon=\pm 1, \label{T-unitary-epsilon-b} 
\end{equation}
  which disentangled form  is
\begin{equation}
\hat{T}_\epsilon = \exp\left[ -  \epsilon \sqrt{\frac{1 -\epsilon \gamma_3}{1+ \epsilon \gamma_3}}  e^{-i (\theta_\alpha - \theta_\beta)} \hat{J}_+ \right]
\exp \left[ \ln\left(   \frac{2}{1 + \epsilon \gamma_3}     \right) \hat{J}_{3}  \right]
\exp\left[   \epsilon \sqrt{\frac{1 -\epsilon \gamma_3}{1+ \epsilon \gamma_3}}  e^{i ( \theta_\alpha - \theta_\beta)} \hat{J}_- \right], \quad \epsilon=\pm 1,
\end{equation} and whose action on the one mode annihilation operators is given by

\begin{equation}
\hat{T}_{\epsilon}^\dagger  \hat{a}_1 \hat{T}_\epsilon = \sqrt{ \frac{1 +\epsilon \gamma_3}{2}} \hat{a}_1 - \epsilon e^{-i ( \theta_\alpha - \theta_\beta )} \sqrt{ \frac{1 - \epsilon \gamma_3}{2}} \hat{a}_2
\end{equation}
and
\begin{equation}
\hat{T}_{\epsilon}^\dagger \hat{a}_2 \hat{T}_\epsilon = \sqrt{ \frac{1 +\epsilon \gamma_3}{2}} \hat{a}_2 + \epsilon e^{i (\theta_\alpha - \theta_\beta)} \sqrt{ \frac{1 - \epsilon \gamma_3}{2}} \hat{a}_1,
\end{equation}
then $\hat{T}_1^\dagger  \hat{H}_{\alpha \beta} \hat{T}_1 = \hbar \;  \hat{a}_1^\dagger \hat a_{1} $  and $\hat{T}_{-1}^\dagger  \hat{H}_{\alpha \beta} \hat{T}_{-1}= \hbar \; \hat{a}_2^\dagger \hat a_{2}.$
 
 \section{Time evolution of the coherent states linked to a class of 2 D  time-dependent isotropic harmonic oscillator}
 \label{sec-five}

Let us now consider the 2 dimensional time-dependent  isotropic harmonic oscillator system governed by the Hamiltonian

\begin{equation}
\hat{H}_{\alpha \beta} (t) =  \hbar  \hat{A}^\dagger_{\alpha\beta}  (t)  \hat{A}_{\alpha \beta} (t),  \label{two-mode-attached-hamiltonian-time-dependent}
\end{equation}
where the  two mode annihilation operator is given by
\begin{equation}
\hat{A}_{\alpha \beta} (t)= \alpha (t) \hat{a}_1  + \beta (t) \hat{a}_2, \label{A-alpha-beta-t}
\end{equation}
where $ \alpha (t)$ and $\beta (t) $ are complex functions which verify
\begin{equation}
\|\alpha (t)\|^2   + \|\beta (t)\|^2 =1, \quad \forall \; t.
\end{equation}
Due to this last fact, the operator  $\hat{A}_{\alpha \beta} (t)$ and its adjoint $\hat{A}^\dagger_{\alpha \beta} (t)$ satisfy the commutation relation
\begin{equation} 
\left[ \hat{A}_{\alpha \beta}  (t), \hat{A}^\dagger_{\alpha \beta}  (t) \right]=\hat{I},
\end{equation} 
where $\hat{I}$ is the identity operator. The explicit form of the time-dependent Hamiltonian (\ref{two-mode-attached-hamiltonian-time-dependent}) is given by

\begin{equation}
\hat{H}_{\alpha \beta}  (s) =  \hbar [ \|\alpha (s) \|^2 \hat{a}^\dagger_1\hat{a}_{1}  +
\|\beta (s) \|^2  \hat{a}^\dagger_2 \hat{a}_{2} + \alpha^\ast (s) \beta (s) \hat{a}^\dagger_1 \hat{a}_2  + \alpha (s) \beta^\ast(s)  \hat{a}^\dagger_2 \hat{a}_1 ],  \label{two-mode-attached-hamiltonian-time-dependent-explicit}
\end{equation}
and as we have done in the previous section, we can make the following correspondence between the parameters
\begin{equation}
w_{11} (s)= \|\alpha (s)\|^2, \quad  w_{22} (s)= \|\beta (s)\|^2  \quad {\rm and} \quad w_{12} (s) = w^\ast_{21}  (s) = \alpha^\ast (s) \beta (s),  \label{parameters-correspondace}
\end{equation} 
or if we define

\begin{equation}
\alpha (s) = \cos ( \rho (s) ) e^{i \theta_{\alpha} (s)} \quad {\rm and} \quad \beta (s) =\sin (\rho (s)) e^{i \theta_{\beta} (s) }, \label{polar-form-alpha-beta}
\end{equation}
we get the following description of them
\begin{equation}
w_{11} (s)=   \cos^2 (\rho (s))  , \quad  w_{22} (s)= \sin^2 (\rho (s))  \quad {\rm and} \quad w_{12} (s) = w^\ast_{21}  (s) =   \cos ( \rho (s)) \sin (\rho (s)) e^{i ( \theta_{\beta} (s) - \theta_{\alpha} (s))}  \label{parameters-correspondace-trigonometric},
\end{equation} 
where $\rho (s), \theta_{\alpha} (s)$ and $\theta_{\beta} (s)$ are time-dependent  real functions. 

Since the set of operators $ \left\{   \hat{A}_{\alpha \beta} (t_0), \hat{A}^\dagger_{\alpha \beta} (t_0), \hat{I},  \hat{H}_{\alpha \beta} (t_0) \right\},  \; \forall t_0 \geq 0,$ generate the one dimensional oscillator algebra, at the instant $t_0$ we can associate to this system  one parameter Schr\"odinger type  coherent states

\begin{equation}
\mid Z_{(t_0)} \rangle  = \exp\left[Z_{ (t_0)}  \hat{A}^\dagger_{\alpha \beta} (t_0) -  Z^\ast_{ (t_0)}  \hat{A}_{\alpha \beta} (t_0) \right] \mid 0,0 \rangle. \label{coherent-states-Z-2-D-t=0}
\end{equation}
Thus, the time evolution of these coherent states, according to   (\ref{coherent-states-Z-2-D-t=0} ), is given by
\begin{equation}
\mid Z_{(t_0)} (t) \rangle =   e^{- \frac{i}{\hbar} \int_{t_0}^{t} \left[ \hat{H}_{\alpha \beta}  (s) \right]_+}  \mid  Z_{ (t_0)} \rangle, \label{time-evolution-coherent-state}
 \end{equation}
where in general $\left[ \hat{H}_{\alpha \beta}  (s) , \hat{H}_{\alpha \beta}  (\tau) \right] \neq 0 $ when $s \neq \tau.$
 
We note that in principle,  $\mid Z_{(0)} (t_0) \rangle \neq \mid Z_{(t_0)} \rangle, $ or  more generally,  $\mid Z_{(t_0)} (t) \rangle \neq \mid Z_{(t)} \rangle.$ When $t_0 =0,$ we can write  $\mid Z_0 \rangle$ in place of   $\mid Z_{(0)} \rangle$ and $\mid Z_{0} (t) \rangle$ in place of $\mid Z_{(0)} (t).$ In the following subsections, without loos of generality, we will suppose that $t_0 =0.$

\subsection{A special case} Let us now consider the special case in which the above parameters verify the relations

\begin{equation}
\|w_{12} (s) \| = \epsilon  \frac{ \eta_0 }{w_0} \frac{d\phi}{ds} (s),  \label{norm-w12-condition} 
\end{equation}
where   $\epsilon = 1 $ or $\epsilon =-1 $ in the region where   $ \frac{d\phi}{ds} (s) > 0$ or  $\frac{d\phi}{ds} (s) <  0, $ respectively,
and
 
 \begin{equation}
\theta_{12} (s) = - \theta_{21} (s) =  \phi (s )+ \frac{\pi}{ 2} - \int_{0}^{s} (w_{11} (\tau) - w_{22}(\tau)) d\tau. \label{phase-12-time-dependent}
\end{equation}
Under these conditions, as we have seen above, the unitary operator present in  (\ref{time-evolution-coherent-state}) can be, in principle, completely disentangled. 
By combining equation (\ref{parameters-correspondace-trigonometric}) with  (\ref{norm-w12-condition}) and  (\ref{phase-12-time-dependent}), we find the following constraints on  the parameters:

\begin{equation}
\frac{d\phi}{ds} (s) = \frac{w_0}{ 2 \; \eta_0} \sin( 2 \rho (s) ) 
\end{equation}
and

\begin{equation}
\theta_{\beta} (s) - \theta_{\alpha} (s) = \phi (s) +  \epsilon \frac{\pi}{2} - \int_{0}^{s} \cos ( 2 \rho (\tau)) d\tau,
\end{equation}
where we have used that $\theta_{12} (s) = \theta_{\beta} (s) - \theta_{\alpha} (s) + (1- \epsilon) \frac{\pi}{2}. $  We note that by integrating the first of these two last equations and performing the derivative of the second  we get

\begin{equation}
\phi (t) = \frac{w_0}{ 2 \; \eta_0} \int_{0}^{t} \sin( 2 \rho (s) ) ds + \phi (0)  \label{phi-t-2-D-time-evolution}
\end{equation}
and
\begin{equation}
\frac{d \theta_{\beta}}{ds}   (s) - \frac{d \theta_{\alpha}}{ds} (s) = \frac{w_0}{ 2 \; \eta_0} \sin( 2 \rho (s) ) -  \cos ( 2 \rho (s)),\label{alpha-beta-t-2-D-time-evolution}
\end{equation}
 which, although it represents an important restriction on the choice of parameters,  it also allows  us some freedom to accommodate the relative phases.

Thus,  by taking into account all these  results at the moment of computing  equation   (\ref{S-matrix-elements-phi-derivative}) we get

\begin{eqnarray}
S_{11} (t) & =&   e^{- \frac{it}{2}}   e^{ \frac{i}{2} \tilde{\Theta}_{\beta \alpha} (t)  } \;  
 \;   \left[ \cos\left( \tilde{\Phi} (t) \right) 
- i    \frac{w_0}{\delta} \; \sin\left(    \tilde{\Phi} (t)    \right) \right] 
\nonumber
 \\ S_{12} (t)  & =&   \;  \frac{  2 \;  \epsilon \; \eta_0}{ \delta} e^{- \frac{it}{2}}  e^{i \phi (0)} \;
 e^{ \frac{i}{2} \tilde{\Theta}_{\beta \alpha} (t) } \;   \sin\left(\tilde{\Phi} (t) \right),    \nonumber \\   S_{21} (t) &=&  -  \;   \frac{ 2 \; \epsilon \;  \eta_0}{ \delta} 
  e^{- \frac{it}{2}}  e^{- i \phi (0)} \; e^{ - \frac{i}{2} \tilde{\Theta}_{\beta \alpha} (t)   } \;   \sin\left( \tilde{\Phi} (t) \right)
 ,  \nonumber\\
 S_{22} (t) & =&    e^{- \frac{it}{2}}   e^{ - \frac{i}{2}  \tilde{\Theta}_{\beta \alpha} (t)   } \;  
 \;  
 \left[ \cos\left( \tilde{\Phi} (t) \right) 
+ i    \frac{w_0}{\delta} \; \sin\left(    \tilde{\Phi} (t)        \right) \right] , 
\nonumber \\ \label{S-matrix-elements-phi-derivative-2-D-time-dependent}
\end{eqnarray} 
where $\phi(0) = \theta_{\beta} (0) - \theta_{\alpha} (0) - \epsilon \frac{\pi}{2},$

\begin{equation}
\tilde{\Phi} (t) = \frac{\delta \tilde{\phi} (t) }{2 w_0}  =   \frac{\delta}{ 4 \eta_{0}}       \int_{0}^{t}  \sin( 2 \rho (s))  ds  \label{phase-tilde-Phi-t}
\end{equation}
and
\begin{equation}
\tilde{\Theta}_{\beta \alpha} (t) = \tilde{\theta}_{\beta} (t) - \tilde{\alpha}_{\alpha} (t)
 = \int_{0}^{t}   \left[ \frac{w_0}{ 2 \; \eta_0} \sin( 2 \rho (\tau) ) -  \cos ( 2 \rho (\tau)) \right] d\tau,    \label{phase-tilde-Theta-t}
\end{equation}
where  $\tilde{\theta}_{\alpha} (t) = \theta_{\alpha} (t) - \theta_{\alpha} (0)$ and  $\tilde{\theta}_{\beta} (t) = \theta_{\alpha} (t) - \theta_{\beta} (0).$

 Hence, with these last expressions in  possession we can compute the time evolution of the coherent states (\ref{coherent-states-Z-2-D-t=0}). In fact, they are given by

\begin{equation}
\mid Z_0 (t) \rangle = \exp\left[ \sum_{\sigma =1}^{2}  c_{\sigma} (t) \hat{a}^\dagger_\sigma - c^\ast_{\sigma} (t) \hat{a}_\sigma \right] \mid 0 , 0 \rangle 
\end{equation}
where 
\begin{eqnarray}
c_1 (t) = S_{11} (t) \tilde{c}_1 (0) + S_{12} (t) \tilde{c}_2 (0) \\
c_2 (t) = S_{21}  (t) \tilde{c}_1 (0) + S_{22} (t) \tilde{c}_2 (0) ,
\end{eqnarray}
which must  verify the initial conditions $c_1 (0) = Z_0 \cos(\rho (0)) e^{-i \theta_{\alpha} (0)} $ and $c_2 (0) = Z_0 \sin(\rho (0)) e^{-i \theta_{\beta} (0)} .$ As according to (\ref{S-matrix-elements-phi-derivative-2-D-time-dependent}), $S_{11}(0)=S_{22}(0)=1$ and $S_{12}(0)=S_{21}(0)=0,$ then from these last two equations we have $\tilde{c}_1 (0) = Z_0 \cos(\rho (0)) e^{-i \theta_{\alpha} (0)} $ and $\tilde{c}_2 (0) = Z_0 \sin(\rho (0)) e^{-i \theta_{\beta} (0)} .$

 Therefore
\begin{eqnarray}
c_1 (t) = S_{11} (t)  Z_0 \cos(\rho (0)) e^{-i \theta_{\alpha} (0)}     + S_{12} (t)   Z_0 \sin(\rho (0)) e^{-i \theta_{\beta} (0)}   \\
c_2 (t) = S_{21}  (t)  Z_0 \cos(\rho (0)) e^{-i \theta_{\alpha} (0)}    + S_{22} (t)  Z_0 \sin(\rho (0)) e^{-i \theta_{\beta} (0)},
\end{eqnarray}
from where

\begin{eqnarray}
c_1 (t) &=&  e^{- \frac{it}{2}}   e^{ \frac{i}{2} \tilde{\Theta}_{\beta \alpha} (t)  } \;  Z_0 \cos (\rho (0))  e^{-i \theta_{\alpha} (0) }  \nonumber \\ 
&\times & \left[ \cos\left( \tilde{\Phi} (t) \right) 
- i    \frac{w_0}{\delta} \; \left(  1  + \frac{2 \eta_{0}}{w_0}  \tan (\rho (0))   \right)   \; \sin\left(    \tilde{\Phi} (t)    \right) \right]  \label{c1-t-2-D}
\end{eqnarray}
and
\begin{eqnarray}
c_2 (t) &=&  e^{- \frac{it}{2}}   e^{ - \frac{i}{2}  \tilde{\Theta}_{\beta \alpha} (t) } \;  Z_0 \sin (\rho (0))  e^{-i \theta_{\beta} (0) } \nonumber \\
& \times & \left[ \cos\left( \tilde{\Phi} (t) \right) 
+ i    \frac{w_0}{\delta} \; \left(  1  -  \frac{2 \eta_{0}}{w_0}  \cot (\rho (0))   \right)   \; \sin\left( \tilde{\Phi} (t) \right) \right]. \label{c2-t-2-D} 
\end{eqnarray}

Finally, inserting these last  coefficients in  (\ref{general-coherent-states}) (for N=2), we obtain the time evolution of the  coherent states associated to the 2 D time-independent isotropic  quantum oscillator Hamiltonian under the action of a special class of extended time-dependent 2 D isotropic quantum oscillator Hamiltonian:

\begin{equation}
\mid Z_0 (t) \rangle  = \exp\left[   c_1 (t) \hat{a}^\dagger_1 -  c^\ast_1 (t) \hat{a}_1 \right]  \exp\left[   c_2 (t) \hat{a}^\dagger_2 -  c^\ast_2  (t) \hat{a}_2 \right]  \mid  0,0 \rangle.
\end{equation}
We note that these coherent states are eigenstates of $\hat{A}_{\alpha \beta} (0) = \cos(\rho (0)) e^{i \theta_{\alpha} (0)} \hat{a}_1 + \sin (\rho (0)) e^{i \theta_{\beta} (0)} \hat{a}_2.$ Indeed, its action on the time-dependent coherent states is given by

\begin{eqnarray}
\hat{A}_{\alpha \beta} (0) \mid Z_0 (t) \rangle &=& \left[ c_{1} (t) \cos (\rho(0)) e^{i \theta_{\alpha} (0)}  + c_{2} (t) \sin (\rho(0)) e^{i  \theta_{\beta} (0)} \right] \; \mid  Z_0 (t) \rangle   \nonumber \\
&= & Z_0  e^{ - \frac{it}{2}} \; \left\{ 
\left[  \cos\left( \frac{  \tilde{\Theta}_{\beta \alpha} (t) }{2} \right) \cos\left( \tilde{\Phi} (t) \right) + \frac{w_0}{\delta}
\sin\left( \frac{  \tilde{\Theta}_{\beta \alpha} (t) }{2} \right) \sin\left( \tilde{\Phi} (t) \right) \right] \right. 
\nonumber \\ &+& i 
\left[  \sin\left( \frac{  \tilde{\Theta}_{\beta \alpha} (t) }{2} \right) \cos\left( \tilde{\Phi} (t) \right) \cos (2 \rho(0)) - \frac{w_0}{\delta}
\left(   \cos (2 \rho(0))    + \frac{2 \eta_0}{ w_0} \sin (2 \rho(0)) \right)  \right.  \nonumber \\  &\times&  \left. \left. \cos\left( \frac{  \tilde{\Theta}_{\beta \alpha} (t) }{2} \right) \sin\left( \tilde{\Phi} (t) \right) \right]  \right\} \;  \mid Z_0(t) \rangle.\end{eqnarray}
On the other hand,  we also have that these coherent states are eigenstates of the generalized time-dependent  lowering operator  
\begin{equation}
\hat{A}_{\left(\frac{ c^\ast_{1}}{Z_0^\ast}  \;  \frac{ c^\ast_{2}}{Z_0^\ast}  \right)}   (t) = \frac{ c^\ast_{1} (t)}{Z_0^\ast}  \hat{a}_1 +   \frac{ c^\ast_{2} (t)}{Z_0^\ast}    \hat{a}_2    
\end{equation}
with  associated eigenvalue equal to $Z_0 .$ Indeed, using the  properties of the  $\hat{S}$ matrix elements,  the definition of the $c$`s coefficients in terms of them and the initial conditions that these coefficients verify, or  calculating  directly  from  (\ref{c1-c2-t-2-D}) and (\ref{c2-t-2-D}), we deduce  that $\|c_{1} (t)\|^2 + \| c_{2} (t)\|^2 = \|Z_0\|^2,$ then 

\begin{equation}
\hat{A}_{\left(\frac{ c^\ast_{1}}{Z_0^\ast}  \;  \frac{ c^\ast_{2}}{Z_0^\ast} \right)}  (t) \mid Z_0(t) \rangle =   \left( \frac{ c^\ast_{1} (t)}{Z_0^\ast} c_{1} (t) +
 \frac{ c^\ast_{2} (t)}{Z_0^\ast} c_2 (t) \right) \mid Z_0(t) \rangle= Z_0 \mid Z_0(t) \rangle.
\end{equation}   
Let us notice that we would have reached the same conclusion if we had worked on Heinsenberg`s picture. 

Moreover, the action of the ladder operator (\ref{A-alpha-beta-t})  on the coherent states $\mid Z_0(t) \rangle$ is given by
 \begin{equation}
\hat{A}_{\alpha \beta} (t)  \mid Z_0 (t) \rangle = \tilde{Z}_{(t)}  \mid Z_0(t) \rangle,
\end{equation}
where

\begin{eqnarray}
 \tilde{Z}_{(t)}  &=&  Z_0 e^{- \frac{it}{2} } e^{\frac{i}{2} \left(\tilde{\alpha} (t) + \tilde{\beta} (t)\right)} \nonumber \\
& \times &
\left\{ \cos\left(   \tilde{\Phi} (t)  \right)  \cos\left(\rho (t) - \rho (0) \right) \nonumber  \right. \\&-&  \left.  \frac{i w_0}{\delta}   \left[     \cos\left( \rho (0) + \rho (t) \right)  + \frac{2 \eta_0}{w_0}  \sin\left(\rho (0) + \rho (t) \right)  \right]   \sin\left(   \tilde{\Phi} (t)  \right)  \right\},
\end{eqnarray}
which can be expressed in a simpler way if we  choose $\tilde{\beta} (t) = - \tilde{\alpha} (t).$

\subsubsection{Case $\rho (s)= \rho_{0},$ linear phases}
When $\rho (s)= \rho_{0},$  where $ \frac{\pi}{2} > \rho_0 > 0, $ from (\ref{phase-tilde-Phi-t}) and (\ref{phase-tilde-Theta-t}) we deduce

 \begin{equation}
\tilde{\Phi} (t) =    \frac{\delta}{ 4 \eta_{0}}   \sin( 2 \rho_0) \; t   
\end{equation}
and
\begin{equation}
\tilde{\Theta}_{\beta \alpha} (t) = \tilde{\theta}_{\beta} (t) - \tilde{\alpha}_{\alpha} (t)
 =  \left[ \frac{w_0}{ 2 \; \eta_0} \sin( 2 \rho_0 )   -  \cos ( 2 \rho_0) \right]  \; t.     
\end{equation}

 Assuming that $\tilde{\theta}_{\alpha} (t) =  - \tilde{\beta}_{\alpha} (t),$ then the  two-mode annihilator (\ref{A-alpha-beta-t}) takes the form

\begin{eqnarray}
\hat{A}_{\alpha \beta} (t) &=&  \cos(\rho_0) \exp\left[i \theta_{\alpha} (0) - \frac{i t}{2} \left(    \frac{w_0}{ 2 \; \eta_0} \sin( 2 \rho_0 ) -  \cos ( 2 \rho_0)        \right)    \right]  \; \hat{a}_{1}  \nonumber \\ &+&  \sin(\rho_0)      \exp\left[i \theta_{\beta} (0 ) + \frac{i t}{2} \left(    \frac{w_0}{ 2 \; \eta_0} \sin( 2 \rho_0 )   -  \cos ( 2 \rho_0)                   \right)    \right]   \;      \hat{a}_{2} ,
\end{eqnarray}
which evaluated in $t=0$ is equal to $\hat{A}_{\alpha \beta} (0) = \cos(\rho_0)  e^{i \theta_{\alpha} (0)} \;  \hat{a}_{1} + \sin(\rho_0)  e^{i \theta_{\beta} (0)} \; \hat{a}_{2}. $ Then the time-evolution of the coherent states $\mid Z_0  \rangle = \exp[Z_0  \hat{A}^\dagger_{\alpha \beta} (0) - Z_0^\ast \hat{A}_{\alpha \beta} (0) ] \mid 0,0 \rangle,$ which are eigenstates of  $\hat{A}_{\alpha \beta} (0)$ with associated eigenvalue equal to $Z_0,$ under the action of the time-ordered unitary operator governed by the Hamiltonian $\hat{H}_{\alpha \beta} (s) = \hat{A}^\dagger_{\alpha \beta} (s)\hat{A}_{\alpha \beta} (s),  $ is represented by the time-dependent coherent states  (\ref{general-coherent-states}), where

\begin{eqnarray}
c_1 (t) &=&  Z_0     e^{- \frac{it}{2}} \cos (\rho (0))   \exp\left[ - i \theta_{\alpha} (0)  + \frac{it}{2}  \left( \frac{w_0}{ 2 \; \eta_0} \sin( 2 \rho_0 )   -  \cos ( 2 \rho_0) \right)  \right] \;   \nonumber \\ 
&\times & \left[ \cos\left( \frac{\delta}{ 4 \eta_{0}}   \sin( 2 \rho_0) \; t  \right) 
- i    \frac{w_0}{\delta} \; \left(  1  + \frac{2 \eta_{0}}{w_0}  \tan (\rho_0 )   \right)   \; \sin\left(    \frac{\delta}{ 4 \eta_{0}}   \sin( 2 \rho_0) \; t    \right) \right]   
\end{eqnarray}
and
\begin{eqnarray}
c_2 (t) &=& Z_0  e^{- \frac{it}{2}}  \sin (\rho (0))  
 \exp\left[ - i \theta_{\beta} (0)  - \frac{it}{2}  \left( \frac{w_0}{ 2 \; \eta_0} \sin( 2 \rho_0 )   -  \cos ( 2 \rho_0) \right)  \right] 
\nonumber \\
& \times & \left[ \cos\left( \frac{\delta}{ 4 \eta_{0}}   \sin( 2 \rho_0) \; t  \right) 
+ i    \frac{w_0}{\delta} \; \left(  1  -  \frac{2 \eta_{0}}{w_0}  \cot (\rho_0)  \right)   \; \sin\left( \frac{\delta}{ 4 \eta_{0}}   \sin( 2 \rho_0) \; t           \right) \right].  
\end{eqnarray}

Finally, the action of the operator  (\ref{A-alpha-beta-t}) on the time-dependent coherent states is given by 
\begin{eqnarray}
\hat{A}_{\alpha \beta} (t)  \mid Z_0(t) \rangle &=&  Z_0 e^{- \frac{it}{2} } 
\left\{ \cos\left(      \frac{\delta}{ 4 \eta_{0}}   \sin( 2 \rho_0) \; t  \right)  \nonumber  \right. \\&-&  \left.  \frac{i w_0}{\delta}   \left[     \cos\left( 2 \rho_0 \right)  + \frac{2 \eta_0}{w_0}  \sin\left( 2 \rho_0 \right)  \right]   \sin\left(   \frac{\delta}{ 4 \eta_{0}}   \sin( 2 \rho_0) \; t                          \right)  \right\}  \mid Z_0(t) \rangle.
\end{eqnarray}
 We notice that if we choose $ \frac{2 \eta_0}{w_0} = \tan(2 \rho_0),$ where $\frac{\pi}{4} \geq \rho > 0,$ then  $\hat{A}_{\alpha \beta} (t) = \hat{A}_{\alpha \beta} (0), $ i.e., the time-dependent coherent states $\mid Z_0 (t) \rangle$ are eigenstates of $ \hat{A}_{\alpha \beta} (0)$ with associated eigenvalue equal to $Z_0 e^{-it},$ as we have seen in section \ref{sec-four}.

 \subsubsection{Logarithmic phases} 
Let us now  particularize the results of the previous subsection to the  case where  $\rho (s)= \arctan (t_0  + s), \;  t_0 > 0, s >0. $ Then  using (\ref{polar-form-alpha-beta}) we get
\begin{equation}
\alpha (s) =  \frac {1}{\sqrt{1 + (s + t_0)^2}   } e^{i \theta_{\alpha} (s)} \quad \text{ and} \quad
\beta (s) =  \frac {s+ s_0}{\sqrt{1 + (s + t_0)^2} } e^{i \theta_{\beta} (s)}, 
\end{equation}
from where we deduce

\begin{equation}
w_{11} (s) = \frac{1}{1 + (s + t_0)^2}, \quad w_{22} = \frac{( s + t_0 )^2} {1 + (s + t_0)^2} \quad \text{ and} \quad   \|w_{12} \| =\|w_{21}\| =  \frac{s + t_0 } {1 + (s + t_0)^2}.
\end{equation}
Moreover, from (\ref{phase-tilde-Phi-t}) and (\ref{phase-tilde-Theta-t}) , we can compute the phases 

\begin{equation}
\tilde{\Phi} (s) = \frac{\delta}{4 \eta_0} \ln \left(\frac{1 + ( s + t_0)^2 }{1 +t^2_0} \right)
\end{equation}
and
\begin{equation}
\tilde{\Theta}_{\beta \alpha} (s) = \frac{w_0}{2 \eta_0} \ln \left(\frac{1 + ( s + t_0)^2 }{1 +t^2_0} \right)  +
s + 2 \arctan (t_0) - 2 \arctan (s + t_0).
\end{equation}

Then, again, if we choose $\tilde{\theta}_{\alpha} (s) = - \tilde{\theta}_{\beta} (s),$ the time-dependent ladder operator writes

\begin{eqnarray}
\hat{A}_{\alpha \beta} (s) &=&  \frac{1}{\sqrt{1 + t_0^2}} \exp\left[i \theta_{\alpha} (0) - \frac{i}{2} 
\left(\frac{w_0}{2 \eta_0} \ln \left(\frac{1 + ( s + t_0)^2 }{1 +t^2_0} \right)  +
s + 2 \arctan (t_0) - 2 \arctan (s + t_0) \right) \right]  \; \hat{a}_{1}  \nonumber \\ &+&   \frac{t_0}{\sqrt{1 + t_0^2}}    \exp\left[i \theta_{\beta} (0 ) + 
\frac{i}{2} \left(\frac{w_0}{2 \eta_0} \ln \left(\frac{1 + ( s + t_0)^2 }{1 +t^2_0} \right)  +
s + 2 \arctan (t_0) - 2 \arctan (s + t_0) \right)   \right]   \;      \hat{a}_{2} . \nonumber \\
\end{eqnarray}

The time-dependent coherent states are obtained from  equation  (\ref{general-coherent-states}),
 with the  $c_{\sigma} (t), \sigma=1,2$  given by 
\begin{eqnarray}
c_1(t) &=&e^{- \frac{it}{2}}  \exp\left[i \frac{ t}{2} +   \frac{i\delta}{4 w_0} \ln \left( \frac{ 1+ (t+ t_0)^2}{1+ t_0^2} \right)  -  i \left( \arctan ( t+ t_0) -\arctan(t_0)  \right)                        \right] \;  Z_0  \frac{ 1}{\sqrt{1 + t_0^2}}   e^{- i \theta_{\alpha} (0)}  {\sqrt{1 + t_0^2}}  \nonumber \\ 
&\times & \left[ \cos\left( \frac{\delta}{4 w_0} \ln \left( \frac{ 1+ (t+ t_0)^2}{1+ t_0^2} \right)\right) 
- i    \frac{w_0}{\delta} \; \left(  1  + \frac{2 \eta_{0} \; t_0}{w_0} \right)   \; \sin\left(
          \frac{\delta}{4 w_0} \ln \left( \frac{ 1+ (t+ t_0)^2}{1+ t_0^2} \right) 
\right) \right] \nonumber  \\
\end{eqnarray}
and
\begin{eqnarray}
c_2(t) &=&e^{- \frac{it}{2}} \exp\left[  - i\frac{ t}{2}  -  \frac{i\delta}{4 w_0} \ln \left( \frac{ 1+ (t+ t_0)^2}{1+ t_0^2} \right)  +  i \left( \arctan ( t+ t_0) -\arctan(t_0)  \right)                        \right] \;  Z_0   \frac{ t_0}{\sqrt{1 + t_0^2}}  e^{- i \theta_{\beta} (0)}  \nonumber \\ 
&\times & \left[ \cos\left( \frac{\delta}{4 w_0} \ln \left( \frac{ 1+ (t+ t_0)^2}{1+ t_0^2} \right)\right) 
+ i    \frac{w_0}{\delta} \; \left(  1  -  \frac{2 \eta_{0} }{w_0 \; t_0} \right)   \; \sin\left(
          \frac{\delta}{4 w_0} \ln \left( \frac{ 1+ (t+ t_0)^2}{1+ t_0^2} \right) 
\right) \right],  \nonumber \\
\end{eqnarray}
where $ \theta_{\beta} (0)$ and $\theta_{\alpha} (0)$ are linked by the relation $\theta_{\beta }(0) = \theta_{\alpha} (0) + \phi(0)  + \frac{\pi}{2} .$
\section{Another equivalent  disentangling order}
\label{sec-six}
   Another equivalent  disentangling order for the unitary operator (\ref{U-t-0-expression}) is,  first isolating an exponential factor raised to a factor proportional to $\hat{N},$  then continuing  with an exponential factor raised to  a factor which is  proportional  the raising  operator $\hat{J}_+,$ to continue with an exponential operator raised to  a factor which is  proportional  to  $\hat{J}_3,$ and   finally leave an exponential factor raised to a factor which is proportional to the lowering operator $\hat{J}_-$   absolutely to the right.  Although, as we have already mention in previous section, by following a new order of disentangling of the operators  we will  obtain the same results, the equation we have to solve to reach them appear in a slightly different form, which could allow to visualize  more easily the possible  resolution methods.   

Then applying the Feynman's disentangling rule in that order we have 
\begin{equation} \hat{U}_0  (t,0) = \exp \left[ - \;  i \; \int_{0}^{t} (w_{11} (s) + w_{22} (s)) ds  \hat{N} \right]
\exp \left [   \tilde{\Lambda} (t) \hat{J}_+  \right] \exp \left [   \tilde{\Omega} (t) \hat{J}_3  \right] \exp \left [   \tilde{\Gamma} (t) \hat{J}_-  \right],
\end{equation}
where $\tilde{\Lambda} (s) , \tilde{\Omega} (s)$ and $\tilde{\Gamma} $(s) verify

\begin{equation}
i \frac{d \tilde{\Lambda}}{ds} (s) =w_{12} (s) + ( w_{11} (s) -w_{22} (s))  \tilde{\Lambda} (s) - w_{21} (s) \tilde{\Lambda}^2 (s), \quad \tilde{\Lambda} (0) =0,  \label{eq-tilde-Lambda}
\end{equation}
\begin{equation}
i \frac{d \tilde{\Omega}}{ds} (s) = ( w_{11} (s) -w_{22} (s)) - 2   w_{21} (s)  \tilde{\Lambda} (s), \quad \tilde{\Omega} (0) =0, \label{Omega-diff-equation}
\end{equation}
 and
 
 \begin{equation}
 \tilde{\Gamma} (s) = - i \int_{0}^{s} w_{21} (\tau) e^{\tilde{\Omega} (\tau)} \; d\tau, \label{Gamma-Omega} 
 \end{equation}
 respectively.

\vspace{1.0cm}

We note that by making $ \tilde{\Lambda} (s) = \exp\left[ -i \int_{0}^{s}  (w_{11} (\tau) - w_{22} (\tau) ) d\tau \right] \Lambda (s), $  equations (\ref{Omega-diff-equation}) and  (\ref{eq-tilde-Lambda})  become
\begin{equation}
i \frac{d \tilde{\Omega}}{ds} (s) = ( w_{11} (s) -w_{22} (s)) + 2 i  \eta^{\ast} (s) \Lambda (s), \quad \tilde{\Omega} (0) =0, \label{eq-Omega-Lambda}
\end{equation}
 and
\begin{equation}
\frac{d \Lambda}{ds} (s) = \eta (s) + \eta^{\ast} (s) \Lambda^2 (s), \quad \Lambda (0) =0,  \label{Lambda-diff-equation-2}
\end{equation}
respectively, where $\eta (s) = - i w_{12} (s) e^{i  \int_{0}^{s}  (w_{11} (\tau) - w_{22} (\tau) ) d\tau},   $ i.e., we have to solve  the same differential equation as in the previous sections.  

\vspace{1.0cm}
Also we observe that (\ref{Lambda-diff-equation-2}) can be transformed into a ordinary second order linear differential equation by making
\begin{equation}\Lambda (s) = - \frac{\dot{u} (s)}{u(s) \eta^{\ast} (s)}, \label{Lambda-change-variable} \end{equation} that is,

\begin{equation}
\ddot{u} (s) - \frac{ \dot{\eta^\ast} (s)}{\eta^{\ast} (s)} \dot{u} (s) + \| \eta (s)\|^2 u(s) =0, \quad \dot{u}(0) =0, \label{u-diff-equation}
\end{equation} 
where $\dot{u}(s) \equiv \frac{du}{ds} (s),$ etc.
On the other hand, by inserting (\ref{Lambda-change-variable}) into equation (\ref{eq-Omega-Lambda}) and then integrating, we get

\begin{equation}
\tilde{\Omega} (s) = - 2 \ln u(s) - i \int_{0}^{s} (w_{11}(\tau) - w_{22} (\tau)) d\tau.
\end{equation} 
In the same way, inserting this last expression into equation (\ref{Gamma-Omega}),  after a little handling of the equations, we get
\begin{equation}
\tilde{\Gamma} (s) = - \int_{0}^{s} \frac{\eta^\ast (\tau)}{u^2 (\tau)} d\tau.
\end{equation}

For example, if $\eta (s) = \eta_0 e^{- i \theta_0 s^2},$ where $\eta_0  $ and $\theta_0$ are real constants verifying $\eta_0 >0,$ 
this last equation takes the form

\begin{equation}
\ddot{u} (s) - 2 i \theta_0 s \dot{u} (s)  + \eta_0^2  u(s) =0, \quad   \dot{u} (0)=0,
\end{equation} 
which solution in terms of the Kummer confluent hypergeometric function is

\begin{equation}
u(s) = {}_1 F_{1} [ i \frac{\eta_0^2}{4 \theta_0}, \frac{1}{2}, i \theta_0 s^2].
\end{equation} 
As 

\begin{equation}
\dot{u} (s) = - \eta_{0}^2 \; s \;  {}_1 F_{1} [1 +  i \frac{\eta_0^2}{4 \theta_0}, \frac{3}{2}, i \theta_0 s^2],
\end{equation}
from the above equations, going back to the original functions we finally get
\begin{equation}
\tilde{\Lambda} (t) = \eta_0 \; t \; \frac{ {}_1 F_{1} [1 +  i \frac{\eta_0^2}{4 \theta_0}, \frac{3}{2}, i \theta_0 t^2]}{ {}_1 F_{1} [ i \frac{\eta_0^2}{4 \theta_0}, \frac{1}{2}, i \theta_0 t^2]} \exp\left[ -i \left(\int_{0}^{t}  (w_{11} (s) -w_{22} (s)) ds + \theta_0 t^2 \right)\right],
\end{equation}

\begin{equation}
\tilde{\Omega} (t) = - 2 \ln \left( {}_1 F_{1} [ i \frac{\eta_0^2}{4 \theta_0}, \frac{1}{2}, i \theta_0 t^2] \right)  - i \int_{0}^{t} (w_{11}(s) - w_{22} (s)) ds
\end{equation} 
and 
\begin{equation}
 \tilde{\Gamma} (t) = -  \eta_0  \int_{0}^{t} \frac{e^{i \theta_0 s^2} \; ds }{\left({}_1 F_{1} [ i \frac{\eta_0^2}{4 \theta_0}, \frac{1}{2}, i \theta_0 s^2]       \right)^2} . \label{tilde-Gamma-quadratic}
 \end{equation}
We note that, at first sight, a closed expression for the integral in (\ref{tilde-Gamma-quadratic}) seems not to be easy to perform, then at first instance, an expansion of  it in Taylor series would be recommended.

\subsection{One more transformation}
We observe that the second  order linear differential equation  (\ref{u-diff-equation}) can be expressed as a first order differential equation system by making the change of variable $ \dot{u} (s) = \eta^\ast (s) v(s), $ where $v(s)$ is an arbitrary function of $s$ which verify $v(0)=0,$ i.e., 

\begin{equation}
\dot{u} (s) - \eta^\ast (s) v(s) =0  \quad \text{and} \quad   \dot{v} (s) + \eta (s) u(s) =0, \quad \dot{u} (0) = v(0)=0.
\end{equation}
Thus, by comparing these last relations, assuming that $\lim_{s \; \rightarrow \;  0} \frac{\dot{u} (s)}{v(s)} $ exist and is equal to $\eta^\ast (0),$ we can write
\begin{equation}
\frac{\dot{v}(s)}{u(s)} = - \frac{\dot{u}^\ast (s)}{v^\ast(s)}, \quad \leftrightarrow \quad u(s) \dot{u}^{\ast} (s)+ \dot{v} (s) v^{\ast} (s)=0. 
\end{equation}
Then, if we express the functions $u(s)$ and $v(s)$ in the polar form, i.e.,  $u(s) = u_0 (s) e^{i \theta_u (s)}$
and  $v(s) = v_0 (s) e^{i \theta_v (s)},$ and insert them into this last equation,  then by separating the real and imaginary part of the resulting equation, we get the following system of differential equations:

\begin{equation}
\frac{1}{2} \frac{d}{ds}\left( u_0^2 (s) + v_0^2 (s) \right) =0, \quad \text{and} \quad v_0^2 (s) \dot{\theta_{v}}(s) - u_0^2 (s) \dot{\theta_{u}} (s)=0. \label{eq-phases-u-v}
\end{equation}
Thus, by integrating the first of these equations, and taking into account the initial conditions, we obtain
\begin{equation}
u_0 (s) = \mid \cos [\varrho (s)] \mid \quad \text{and} \quad v_0 (s) = \mid \sin [\varrho (s)] \mid, \quad \varrho (0)=0,  \label{u0-v0-cos-sin}
\end{equation}
and then the second of these equations becomes

\begin{equation}
\dot{\theta}_{u} (s)= \tan^2 [\varrho (s)] \dot{\theta}_{v} (s), \label{eq-theta-u-v-varrho}
\end{equation}
where we must add the constraint 

\begin{equation}
\eta (s) = - \left( \frac{\dot{v}_0 (s)}{u_0 (s)} + i \dot{\theta}_v (s) \frac{v_0 (s)}{u_0 (s)}   \right) \exp\left[ i \left(\theta_v (s) - \theta_u (s) \right) \right]= - i w_{12} (s) e^{i  \int_{0}^{s}  (w_{11} (\tau)- w_{22} (\tau) ) d\tau}, \label{eta-definition-u-v}
\end{equation}
which is separated in two when comparing the norms  and phases:

\begin{equation}
\| w_{12} (s) \|^2 = \dot{\varrho}^2 (s) +   \tan^2 \left( \varrho (s)\right) \dot{\theta}^2_v (s) \label{eq-varrho} \end{equation}
and
\begin{equation}
\arctan \left(\frac{\dot{\theta}_v (s)}{\dot{\varrho} (s)} \tan\left( \varrho (s)  \right)\right) + \theta_{v} (s) - \theta_{u} (s) = \frac{\pi}{2} + \theta_{12} (s) + \int_{0}^{s} (w_{11} (\tau) - w_{22}(\tau)) d\tau, \quad \dot{\varrho} (s) \neq 0. \label{eq-constraint}
\end{equation}

In terms of the new variables we have

\begin{eqnarray}
\tilde{\Lambda} (t) &=& - \frac{v (t)}{u(t)} \exp\left[ - i \int_{0}^{t} (w_{11} (s) - w_{22}(s)) ds  \right]\nonumber \\ &=& - 
\left| \tan \left( (\varrho (t) \right) \right| \exp[i (\theta_v (t) - \theta_u (t))] \exp\left[ - i \int_{0}^{t} (w_{11} (s) - w_{22}(s)) ds\right], \label{eq-L-1}
\end{eqnarray}

\begin{equation}
\tilde{\Omega} (t) =  \ln \left( \sec^2 ({\varrho}(t))  \right) - 2 i \theta_u (t)  - i \int_{0}^{t} (w_{11}(s) - w_{22} (s)) ds, \quad \theta_u (0)= 0,\label{eq-Q-2}
\end{equation} 
and 
\begin{equation}
 \tilde{\Gamma} (t) =  \int_{0}^{t}  \|w_{12} (s)\| \left(1 + \tan^2 \left( \varrho (s)\right)\right)\exp\left[-i \left[
 \arctan \left( \frac{\dot{\theta}_v (s)}{\dot{\varrho} (s)} \tan\left( \varrho (s)  \right)\right) + \theta_{v}(s) + \theta_u (s) \right] \right] ds.\label{eq-G-3}
 \end{equation}

\subsubsection{Case $\theta_u (s) =0$ }
When  $\theta_u (s) =0,$ from (\ref{eq-phases-u-v}) we deduce that  $\theta_{v}(s)= \theta_v^0$ is constant. Then, from (\ref{eq-varrho}) we get\footnote{Assuming that $\varrho (s) $ is an increasing  function of $s.$}

\begin{equation}
\dot{\varrho}(s) = \|w_{12} (s)\|, \quad \text{or} \quad  {\varrho}(s) = \int_{0}^{s} \|w_{12} (\tau)\| d\tau.
\end{equation}
Thus, by inserting this last result into equations (\ref{eq-L-1}), (\ref{eq-Q-2}) and (\ref{eq-G-3}), we obtain

\begin{equation}
\tilde{\Lambda} (t) = - 
\left| \tan \left(   \int_{0}^{s} \|w_{12} (s)\| ds  \right) \right| \exp[i \theta_v^0 ] \exp\left[ - i \int_{0}^{t} (w_{11} (s) - w_{22}(s)) ds\right], 
\end{equation}

\begin{equation}
\tilde{\Omega} (t) =  \ln \left( \sec^2 ({\varrho}(t))  \right)  - i \int_{0}^{t} (w_{11}(s) - w_{22} (s)) ds,
\end{equation} 
and
\begin{equation}
 \tilde{\Gamma} (t) = \left| \tan \left[  \int_{0}^{t}  \|w_{12} (s)\|  ds \right]\right| \exp\left[-i \theta_v^0
  \right], 
 \end{equation}
respectively,  where the following constraint must be respected

\begin{equation}
 \theta_{v}^0  = \frac{\pi}{2} + \theta_{12} (s) + \int_{0}^{s} (w_{11} (\tau) - w_{22}(\tau)) d\tau. 
\end{equation}

\subsubsection{Case $\dot{\theta}_v (s) =\dot{\varrho} (s)$ and Fresnel type solutions }
When  $\dot{\theta_v} (s) = \dot{\varrho} (s),$ i.e, when $\theta_v (s) = \varrho (s) + \theta_v^0,$  from (\ref{eq-phases-u-v}) we deduce   \begin{equation}
\dot{\theta}_{u}(s)= \tan^2 \left( \varrho (s) \right) \dot{\varrho} (s),
\end{equation}
which implies that

\begin{equation}
\theta_{u} (s) = \tan \left( \varrho (s)\right) - \varrho (s) + \theta_u^0.
\end{equation}
 Thus, inserting the results into  (\ref{eq-varrho}), we get
 
 \begin{equation}
\dot{\varrho}(s) \sec \left(  \varrho (s) \right) = \pm  \| w_{12} (s) \|, 
\end{equation}
whose solution is given by

\begin{equation}
\varrho (s) = 2 \arctan \left\{ \tanh \left[ \pm \frac{1}{2} \left( \int_{1}^{s}   \| w_{12} (\tau) \| d\tau -   \int_{1}^{0}   \| w_{12} (\tau) \| d\tau \right) \right] \right\}.
\end{equation}

Therefore,  in this case, the constraint (\ref{eq-constraint}) writes 
\begin{equation}
  3 \varrho (s)  - \tan \left( \varrho (s)  \right) + \theta_{v}^0 - \theta_{u}^0 = \frac{\pi}{2} + \theta_{12} (s) + \int_{0}^{s} (w_{11} (\tau) - w_{22}(\tau)) d\tau. \label{restrictions}
\end{equation}

\vspace{1.0cm}
For example, choosing $\| w_{12} (s) \| = w_{12}^0 | \cos \left( \nu s^2 \right)|, $ where $ w_{12}^0 $ and $\nu $ are non-negative real constants, we get
\begin{equation}
\varrho(s) = 2  \arctan  \left\{ \tanh \left[  \frac{ w_{12}^0  \sqrt{\pi}  \sqrt{1 + \cos (2 \nu s^2)} \mathbb{F}_{C} \left(  \sqrt{\nu} \sqrt{\frac{2}{\pi}} s \right)  \sec (\nu s^2)}{4  \sqrt{\nu}}       \right] \right\},
\end{equation}
where
\begin{equation}
\mathbb{F}_{C} (x) = \int_{0}^{x} \cos (\frac{\pi}{2} \tau^2) d\tau = \sqrt{\frac{2}{\pi}}\sum_{n=0}^{\infty} (-1)^n \frac{ (\sqrt{\frac{\pi}{2}} x)^{4n+1}}{(2n)! (4n+1) }, 
\end{equation}
is the Fresnel cosinus integral \cite{AMSI-table}.

Finally, by inserting these  results into equations (\ref{eq-L-1}), (\ref{eq-Q-2}) and (\ref{eq-G-3}), we can obtain the desired disentangling of the unitary operator (\ref{U-t-0-expression}), under the restrictions imposed by equation (\ref{restrictions}). 

\subsubsection{An additional trigonometric reparameterization  }
Looking at equation (\ref{eq-varrho}), we observe that the additional trigonometric reparameterization   \begin{equation}\dot{\varrho} (s) = \| w_{12} (s) \| \cos(\Theta (s)) \quad  \text{and} \quad \tan(\varrho (s) \dot{\theta}_v (s) =  \| w_{12} (s) \| \sin (\Theta (s)), \label{varrho-dot-theta-v-dot}\end{equation} where $\Theta (s)$ is a real function\footnote{If we suppose that $\dot{\varrho} (s) > 0,$ then we can choose  $ - \frac{\pi}{2} < \Theta(s) < \frac{\pi}{2}. $}of $s,$ could help us to better visualize the global structure of the final disentangled form of the unitary time evolution operator. In terms of this new function $\Theta (s),$ equation (\ref{eq-varrho}) verifies automatically, whereas equations  (\ref{eq-theta-u-v-varrho}) and (\ref{eq-constraint}) become
\begin{equation}
\dot{\theta}_{u} (s) = \| w_{12} (s) \| \tan (\varrho (s)) \sin(\Theta (s)), \label{eq-u-big-theta}
\end{equation}
and
\begin{equation}
 \Theta (s) + \theta_{v} (s) - \theta_{u} (s) = \frac{\pi}{2} + \theta_{12} (s) + \int_{0}^{s} (w_{11} (\tau) - w_{22}(\tau)) d\tau, \quad  \dot{\varrho} (s) \neq 0, \label{eq-big-theta-constraint}
\end{equation}
respectively. For example, when $\dot{\varrho} (s) \neq 0,$   the function $\tilde{\Gamma} (t) $   in (\ref{eq-G-3}) assumes the form
 
\begin{equation}
 \tilde{\Gamma} (t) =  \int_{0}^{t}  \|w_{12} (s)\|  \sec^2\left[ \varrho (s)\right] \exp\left[-i \left[ \Theta (s) + \theta_{v}(s) + \theta_u (s) \right] \right] ds,\label{eq-G-3-Theta}
 \end{equation}
 which can be  directly integrated with the help of equations   (\ref{varrho-dot-theta-v-dot} ) and (\ref{eq-u-big-theta}), we get
 \begin{equation}
 \tilde{\Gamma} (t) =  \left| \tan\left( \varrho (t) \right) \right|   \exp\left[- i \left( \theta_{v}(t) + \theta_u (t) \right) \right].\label{eq-G-3-Theta-final}
 \end{equation}

So, proceeding in this way, we can generate disentangled expressions for the original unitary time evolution operator. As we have mentioned before, we cannot solve the original Riccati differential equation for any value of the time-dependent  $w$ functions, nor we can do it with the alternative transformed versions  of this equation.   However, we have proven here that this is possible when there are appropriated connections among these parameters.

\section{Conclusions}
In this article we have used the Feynman disentangling operator rules to disentangle the time evolution operator of a system which Hamiltonian is an   element of the complex  $\left\{h(1) \oplus h(1) \right\} \roplus u(2)$ composed Lie algebra. For such a purpose, we  have taken advantage of the two-boson representation space of this algebra. In particular, we have written some generators in terms of the  generators of the $su(2)$ algebra in the two-boson Schwinger representation.    We have proven that the complete disentangling of that time evolution operator depends exclusively on our capacity to solve  a standard Riccati-type equation with time dependent coefficients, which are related  directly with the time-dependent frequencies, or set of coefficients,  of the $u(2)$ expansion part of the original Hamiltonian.  In other words, in the case $N=2,$ we have proved that the conditions to  disentangle the temporal evolution operator are reduced to determining the appropriate combination of frequency values that allow us to solve the Riccati-type differential equation. All that because we know  that not always is possible to have a closed solution for such non-linear differential equation.  Nevertheless, we have found some relations among the frequencies that allow us to solve that differential equation and disentangle completely the unitary time evolution  operator. By the way, these relations are not the only ones that work. Here, in  section \ref{sec-six}, we have recalled some others that can lead us to solutions in terms of hypergeometric or Fresnel functions, for example.

Another important fact we have illustrated  in this article is that once that the  disentangling of the $u(2)$ sector of the unitary evolution operator has been done, the matrix elements of this  same disentangled factor can be used to complete the  disentangling of the $h(1) \oplus h(1) $ sector,  no matter   the representation we use to compute these matrix elements. That because it have been shown that the differential equation system, which serve to untangle the exponential factor containing the linear terms in the unitary time evolution operator, can be expressed as a time ordered operator equation.  It is in this sense that we say that the Feynman disentanglement method is self-consistent in dealing with this problem.

 By way of example, we have computed the time evolution of a class of coherent states linked to the 2 D isotropic harmonic oscillator as well as to a class of coherent states linked to a system of two time-dependent $u(2)$ interacting oscillators. This study could be of some interest from the fact that here we have shown that we can associate the same set of coherent states to two quantum systems whose Hamiltonians are not connected by a unitary transformation.
 
  \section*{Acknowledgments}
 This article is dedicated to the memory of Dr. Luciano Edmundo Laroze Barrios, teacher at Department of Physics,  Universidad Técnica
  Fedérico Santa María, Valparaíso, Chile:  {\it Teacher, could you be my thesis supervisor? Neither a solemn yes nor a resounding no for an answer, just a big smile and body movements launched into space, imitating the gestures of a marine diver submerged in the ocean and saying out loud, we will dive in time!} July 1988.

  \appendix
\section{Solving a differential equation system by using the concept of  time ordering operator}
\label{appa}
The untangled closed form of the general time evolution operator of the system we are dealing with will not be possible until the solution of the differential equation system  (\ref{eq-sys-differential})  is available. I have been shown in the literature, see  \cite{NEAM-92},  that  (\ref{eq-sys-differential})  can be solved for any number of interacting oscillators using the concept of time ordering operator. Indeed, let us consider a complete orthonormal set of states $ \left\{\mid \psi_\sigma \rangle \right\} $ where the functions $ c_{\sigma} (t),$ $w_{\sigma \lambda} (t)$ and $F_{\sigma} (t)$ be represented by

\begin{equation}
c_{\sigma} (t) = \langle \psi_\sigma \mid  \chi (t) \rangle, \quad w_{\sigma \lambda} (t) =  \langle \psi_\sigma \mid \hat{ W (t)}
(t)  \mid \psi_\lambda \rangle, \quad {\rm and} \quad F_{\sigma} (t) =\langle \psi_\sigma \mid F (t) \rangle,   
 \end{equation}  
where $\hat{W}$ is an appropriate  operator acting on these basis states.  Then, with these definitions (\ref{eq-sys-differential})   assumes the  form

\begin{equation}
i \hbar \frac{d}{dt} \mid \chi (t) \rangle - \hbar \hat{W} (t)  \mid \chi (t) \rangle - \mid F (t) \rangle =0 . \label{eq-bra-ket}
\end{equation} 
The associated homogeneous associated equation of this latter is given by

\begin{equation}
 i  \frac{d}{dt} \mid \Phi (t) \rangle =  \hat{W} (t)  \mid \Phi (t) \rangle,
\end{equation}
which is a  equation of the Schr\"odinger type, whose solution can be found using the time ordering technique, that is

\begin{equation}
 \mid \Phi (t) \rangle = \hat{S} (t,t_0)  \mid \Phi (t_0) \rangle,
\end{equation}   
where
\begin{equation}
\hat{S} (t,t_0) = e^{-i   \int_{t_0}^{t} \left[ \hat{W}  (s) \right]_+ ds} . \label{unitary-operator-solution}
\end{equation}
Now,  if we try in  (\ref{eq-bra-ket}) a particular solution  $\mid \chi_{p} (t) \rangle$ of the type  

\begin{equation}
\mid \chi_{p} (t) \rangle = \hat{S}(t,t_0 ) \mid Z (t) \rangle, 
\end{equation}
and then we  perform some reductions to the resulting expressions, we get

\begin{equation}
i \hbar \frac{d}{dt} \mid Z (t) \rangle = \hat{S}^{\dagger} (t,t_0)  \mid F (t) \rangle,
\end{equation}
from where
 \begin{equation}
 \mid Z (t) \rangle = - \frac{i}{\hbar} \int_{t_0}^{t}  \hat{S}^{\dagger} (s,t_0) \mid F (s) \rangle ds, 
\end{equation}
and by consequence
\begin{equation}
\mid \chi_{p} (t) \rangle=  -    \frac{i}{\hbar}    \hat{S} (t,t_0) \;   \int_{t_0}^{t}  \hat{S}^{\dagger} (s,t_0) \mid F (s) \rangle ds. 
\end{equation}
In this way we obtain the general solution of  (\ref{eq-bra-ket})  in the operator form

\begin{equation}
\mid \chi (t) \rangle = \hat{S} (t,t_0) \mid \chi (t_0) \rangle -  \frac{i}{\hbar} \hat{S} (t,t_0)  \int_{t_0}^{t}  \hat{S}^{\dagger} (s,t_0) \mid F (s) \rangle ds. 
\end{equation}
By  projecting  this  general expression on the basis states $\mid \psi_\sigma \rangle,$ we finally arrive to 

\begin{equation}
c_{\sigma} (t) = \sum_{\lambda} S_{\sigma \lambda} (t,t_0) \tilde{c}_{\lambda} (t_0) - \frac{i}{\hbar} \sum_{\lambda \nu} S_{\sigma \lambda} (t,t_0) \int_{t_0}^{t} S^\ast_{\nu \lambda} F_{\nu} (s) ds, \label{c-solutions}
\end{equation} 
where the superscripts under the sum symbols  run from 1 to the total number of time-dependent oscillators, in our case it is equal to 2. 

\end{document}